%% file: main_eng.tex
\documentclass[aps,prb,10pt,twocolumn,superscriptaddress,a4paper]{revtex4-2}
\usepackage[dvipdfmx]{graphicx}

\usepackage{amsmath,amsthm,amssymb}
\usepackage{bm}
\usepackage{color}
\usepackage{ifthen}
\usepackage[svgnames]{xcolor}
\usepackage{multirow,bigdelim}
\usepackage{comment}

\usepackage[
pagebackref=false,
colorlinks=true,
linkcolor=blue,
urlcolor=blue,
filecolor=black,
citecolor=blue,
pdfstartview=FitV,
pdftitle={},
pdfauthor={},
pdfsubject={},
pdfkeywords={},
bookmarksopen=true
]{hyperref}

\usepackage{url}
\usepackage{orcidlink}

\newcommand{\1}{\mbox{1}\hspace{-0.25em}\mbox{l}}

\def\Tr{\mathrm{Tr}}

\newcommand{\ii}{\mathrm{i}}

\newcommand{\sectitle}[1]{\phantomsection\emph{{#1}.---}\addcontentsline{toc}{section}{#1}\ignorespaces}

\begin{document}

%%%%%%%%%%%%
\title{
Interaction-Enabled Two- and Three-Fold Exceptional Points
}
%%%%%%%%%%%%
\author{
Musashi Kato
}
\email{kato.musashi.52d@st.kyoto-u.ac.jp}
\affiliation{%
Department of Physics, Kyoto University, Kyoto 606-8502, Japan
}%
\author{
Tsuneya Yoshida\orcidlink{0000-0002-2276-1009}
}
\email{yoshida.tsuneya.2z@kyoto-u.ac.jp}
\affiliation{%
Department of Physics, Kyoto University, Kyoto 606-8502, Japan
}%

\date{\today}
\begin{abstract}
We propose a novel type of exceptional points, dubbed interaction-enabled $n$-fold exceptional points [EP$n$s  ($n=2,3$)] ---EP$n$s protected by topology that are prohibited at the non-interacting level.
Specifically, we demonstrate that both bosonic and fermionic systems host such interaction-enabled EP$n$s ($n=2,3$) in parameter space that are protected by charge U(1), pseudo-spin-parity, and $PT$ symmetries. The interaction-enabled EP2s are protected by zero-dimensional topology and give rise to qualitative changes in the loss rate, an experimentally measurable quantity for cold atoms. Furthermore, we reveal that interactions enable EP3s protected by one-dimensional topology beyond the point-gap topological classifications, 
suggesting the potential presence of a broader class of interaction-enabled 
non-Hermitian degeneracies.
\end{abstract}
\maketitle

%%%%%%%%%%%%%%%
\sectitle{
Introduction
}
%%%%%%%%%%%%%%%
Over the past few decades, extensive efforts have revealed that the interplay between many-body interactions and Hermitian topology gives rise to a variety of exotic phenomena that do not emerge in non-interacting systems~\cite{Fidkowski_CorrHTopo_PRB2011,Gurarie_CorrHTopo_PRB2011,Turner_PRB2011,Yao_PRB2013,You_PRB2014,Wang_Sci2014,Morimoto_PRB2015}.
Among them, it has been found that correlations can change topological classifications
that provide a unified understanding of non-interacting topological insulators~\cite{Schnyder_PRB2008,Kitaev_AIP2009,Ryu_NJP2010}. For instance, correlations can reduce the number of possible topological phases; for one-dimensional topological superconductors of symmetry class BDI, the $\mathbb{Z}$ classification in the non-interacting case is reduced to $\mathbb{Z}_8$ due to strong correlations~\cite{Fidkowski_CorrHTopo_PRB2011}.
Furthermore, correlations also enable topological phases that are prohibited at the non-interacting level; for one-dimensional topological superconductor of class BDI with inversion symmetry, the classification result changes from $0$ to $\mathbb{Z}_2$ due to strong correlations~\cite{Lapa_IETI_PRB2016}.
Correlation effects on topological classifications have been explored in higher dimensions~\cite{Yao_PRB2013,You_PRB2014,Wang_Sci2014,Morimoto_PRB2015}
and further generalized to include topology in parameter space~\cite{QNiu_JPhysA1984,Wang_PRL2013,Nakagawa_PRB2018}, as exemplified by the classification of topological pumps under strong correlations~\cite{Jian_PRX2018}. 

While many-body correlations induce rich topological phases in Hermitian systems, non-Hermiticity has opened a parallel avenue in non-interacting systems~\cite{Martinez_PRB2018,Yao_PRL2018,YaoFei_PRL2018,Kunst_PRL2018,Yokomizo_PRL2019,Lee_PRB2019,Borgnia_PRL2020,Zhang_PRL2020,Okuma_PRL2020,Chang_PRRes2020,Kawabata_PRX2023,Gong_PRX2018,Kawabata_PRX2019,Zhou_PRB2019,Ashida_AdPhys2020,Bergholtz_RMP2021}. 
In non-Hermitian systems, complex-valued eigenvalues lead to unique topology known as the point-gap topology~\cite{Gong_PRX2018,Kawabata_PRX2019,Zhou_PRB2019}. The point-gap topology protects 
characteristic degeneracies known as exceptional points (EPs)~\cite{Kato1966,Zhou_Sci2018,Shen_PRL2018,Kawabata_PRL2019,Wojcik_PRB2020,Yang_2021}.
At these points, as well as eigenvalues, two eigenstates coalesce, resulting in 
the dispersion relation with a fractional exponent~\cite{Kozii_PRB2024,Yoshida_PRB2018,Yoshida_PRB2019,Budich_PRB2019,Okugawa_PRB2019}.
This type of unique non-Hermitian degeneracies
is further extended to $n$-fold exceptional points (EP$n$s with $n\geq 3$)~\cite{Delplace_2021,Mandal_PRL2021,Sayyad_PRR2022,Montag_PRR2024,Yoshida_2025}, or higher-order EPs, which are protected by topology 
beyond the point-gap topological classifications~\cite{Delplace_2021,Yoshida_2025}. 
These EPs are realized in a wide variety of systems ranging from classical metamaterials~\cite{Dembowski_PRL2001,Lee_PRL2009,Xiao_NatPhys2017,Chen_Nat2017,Hodaei_Nat2017,Miri_Sci2019,Doppler_Nature2016,Zhang_2019,Ding_PRX2016,Tang_Sci2020,Tang_NatComm2023}
 to open quantum systems~\cite{Naghiloo_NatPhys2019,Chen_PRL2022,Han_NatComm2024,Ding_PRL2021,Lu_CommPhys2025,Chen_NatComm2025,Liu_PRL2021,Wu_NatNano2024,Li_NatComm2019,Ren_NatPhys2022,Wang_PRL2024,Zhao_Nat2025,Zhang_NatComm2025,Chen_PRL2025},
 including NV centers~\cite{Liu_PRL2021,Wu_NatNano2024} and cold atoms~\cite{Li_NatComm2019,Ren_NatPhys2022,Wang_PRL2024,Zhao_Nat2025,Zhang_NatComm2025,Chen_PRL2025}. In particular, for cold atoms, the high controllability allows access to the non-Hermitian topology under correlations in parameter space~\cite{Zhang_NatComm2025,Chen_PRL2025}.
 
The above progress naturally raises the following crucial question: \textit{can correlated EPs emerge without non-interacting counterparts?}
Although correlation effects on topology unique to non-Hermitian systems
have been addressed so far~\cite{McClarty_PRB2019,Mu_PRB2020,Faugno_PRL2022,Shen_CommPhys2022,Zhang_PRB2022,Kawabata_PRB2022,Hamanaka_PRB2023,Yoshida_EPReduction_PRB2023,Yoshida_PRL2024,Kim_CommPhys2024,Ekman_PRR2024,Garcia_PRB2025,Ling_PRB2025,Pan_PRA2019,Luitz_PRR2019,Sch_PRR2022}, interaction-enabled topology remains largely unexplored, especially in the context of $n$-fold EPs with $n\geq 3$.

In this letter, we propose a novel type of non-Hermitian degeneracies,  
dubbed interaction-enabled EP$n$s ($n=2,3$) ---EPs protected by non-Hermitian topology that are 
prohibited at the non-interacting level.
Specifically, analyzing second-quantized Hamiltonians, we demonstrate the emergence of interaction-enabled EP$n$s ($n=2,3$) in parameter space for both bosonic and fermionic systems with charge U(1), pseudo-spin-parity, and $PT$ symmetries. 
The interaction-enabled EP2s are protected by zero-dimensional topology and give rise to a qualitative difference in loss rate, an experimentally measurable quantity for cold atoms.
Furthermore, we demonstrate that interactions enable EP3s protected by one-dimensional topology beyond the point-gap topological classifications, suggesting the potential presence of a broader class of interaction-enabled non-Hermitian degeneracies.

%%%%%%
\sectitle{
Interaction-enabled zero-dimensional point-gap topology
}
%%%%%%
We consider a two-component bosonic system 
in a two-dimensional parameter space $\bm{\lambda}=(x,y)$, where $x$ and $y$ are real parameters. The second-quantized Hamiltonian is written as
%%%%%%
\begin{align}
\hat{H}(\bm{\lambda}) = \hat{H}_0(\bm{\lambda}) + \hat{H}_{\mathrm{int}},
\end{align}
%%%%%%
where $\hat{H}_0$ denotes the quadratic (non-interacting) 
part and
$\hat{H}_{\mathrm{int}}$ represents the many-body interaction.
We impose the following symmetries on $\hat{H}(\bm{\lambda})$:
%%%%%%
\begin{align}
\label{eq: IEEL c-U1 symm}
[\hat{H}(\bm{\lambda}),\hat{N}]&=0, \\
\label{eq: IEEL spin-parity symm}
[\hat{H}(\bm{\lambda}),(-1)^{\hat{T}_z}]&=0, \\
\label{eq: IEEL PT-H}
 \hat{P}\hat{T} \hat{H}(\bm{\lambda}) ( \hat{P}\hat{T})^{-1}
 &= \hat{H}(\bm{\lambda}).
\end{align}
%%%%%%
Here, $\hat{P}\hat{T}$ is an antiunitary operator satisfying $(\hat{P}\hat{T})^2=\1$, where $\1$ denotes the identity operator. It can be represented as the product of a unitary operator and the complex conjugation operator $\mathcal{K}$ ($\mathcal{K}\mathrm{i} = -\mathrm{i}\mathcal{K}$).
The operators $\hat{N}$ and $\hat{T}_z$ denote the total particle-number operator and the pseudo-spin operator, respectively, and are defined as $\hat{N}=\hat{N}_\mathrm{A}+\hat{N}_\mathrm{B}$, $\hat{T}_z=(\hat{N}_{\mathrm{A}}-\hat{N}_{\mathrm{B}})/2$, where $\hat{N}_{\mathrm{A}}$ ($\hat{N}_{\mathrm{B}}$) is the particle-number operator for the $\mathrm{A}$ ($\mathrm{B}$) component. We further assume that these operators satisfy the following relations with the $PT$ operator:
%%%%%%
\begin{align}
 [\hat{P}\hat{T},\hat{N}]&=0,  \\    
\{\hat{P}\hat{T},\hat{T}_z\}&=0.
\end{align}
%%%%%%
Here, for arbitrary operators $\hat{A}$ and $\hat{B}$, the commutation and anticommutation
relations are denoted by 
$[\hat{A},\hat{B}]=\hat{A}\hat{B}-\hat{B}\hat{A}$, 
$\{ \hat{A},\hat{B}\}=\hat{A}\hat{B}+\hat{B}\hat{A}$, respectively.

An analysis of the zero-dimensional point-gap topology reveals that correlations change  
the allowed topological structures:
%%%%%%
\begin{subequations}
\label{eq: boson EP2 class results}
\begin{align}
\ \ 0 \rightarrow\mathbb{Z}_2 \quad\quad &\mathrm{for}\ N+1+\sigma=0 \pmod{4},\\
\mathbb{Z}_2 \rightarrow\mathbb{Z}_2 \quad\quad &\mathrm{for}\ N+1+\sigma=2 \pmod{4},\\
0 \rightarrow 0 \quad\quad &\mathrm{for}\ N+1+\sigma=1,3 \pmod{4},
\end{align}
\end{subequations}
%%%%%%
where $N$ and $\sigma$ denote eigenvalues of $\hat{N}$ and $(-1)^{\hat{N}_{\mathrm{B}}}=(-1)^{\hat{N}/2-\hat{T}_z}$, respectively.
In particular, for $N+1+\sigma=0 \pmod{4}$, interactions enable the nontrivial point-gap topology, which we see in detail below (for other cases, see Sec.~\ref{sec: Fock-sp} of Supplemental Material~\cite{Suppl}).

Firstly, we block-diagonalize the second-quantized Hamiltonian.
In correlated cases,
Eqs.~\eqref{eq: IEEL c-U1 symm} and \eqref{eq: IEEL spin-parity symm}
imply that the Hamiltonian $\hat{H}=\hat{H}_0+\hat{H}_{\mathrm{int}}$ commutes with both $\hat{N}$ and $(-1)^{\hat{N}_{\mathrm{B}}}$, indicating that
$\hat{H}$ can be block-diagonalized in Fock space
into sectors labeled by their eigenvalues $N$ and $\sigma$; $\hat{H}=\bigoplus_{N,\sigma} H_{(N,\sigma)}$.
In the non-interacting case, the Hamiltonian $\hat{H}=\hat{H}_0$
satisfies an additional symmetry $[ \hat{H}_0, \hat{T}_z ] = 0$.
This property follows from the symmetry constraint
[Eq.~\eqref{eq: IEEL spin-parity symm}] and from the fact that
$\hat{H}_0$ is quadratic in the creation and annihilation operators.
Therefore, in the absence of interactions, 
$H_{(N,\sigma)}$
can be further block-diagonalized by $\hat{T}_z$; $H_{(N,\sigma)}=\bigoplus_{T_z} H_{0(N,\sigma,T_z)}$ with $T_z$ being the eigenvalues of $\hat{T}_z$.
Because of the anticommutation relation,
$\{ \hat{P}\hat{T}, \hat{T}_z \} = 0$, 
the $PT$ symmetry is not closed within the sector with a finite value of $T_z$.

Let us focus on the case for $N+1+\sigma = 0 \pmod{4}$.
In the absence of interactions, within this Fock-space sector
$\hat{P}\hat{T}$ and $\hat{H}$ are represented as
%%%%%%
\begin{align}
\label{eq: IEEL PT example}
PT_{(N,\sigma)} &=
\begin{pmatrix}
    O & {p^*}^{-1} \\
    p & O
\end{pmatrix} \mathcal{K}, \\
\label{eq: IEEL free H example}
H_{(N,\sigma)} &=
\begin{pmatrix}
   H_{0+(N,\sigma)} & O \\
    O & H_{0-(N,\sigma)}
\end{pmatrix}.
\end{align}
%%%%%%
Here, $O$ denotes the zero matrix, $p$ is a unitary operator.
The matrices $H_{0+(N,\sigma)}$ and $H_{0-(N,\sigma)}$ represent the Hamiltonian acting on the subspaces
of the Fock-space sector $(N,\sigma)$ with positive ($+$) and negative ($-$)
eigenvalues of $\hat{T}_z$, respectively.
We note that the subspace of $T_z=0$ is not included since $\hat{T}_z$ is odd for $N+1+\sigma = 0 \pmod{4}$.
The zero-dimensional topology of the $H_{(N,\sigma)}$ is characterized by the topological index~\cite{Gong_PRX2018,Kawabata_PRX2019,Kawabata_PRL2019,Kawabata_PRB2022}
%%%%%%
\begin{align}
\label{eq: IEEL Z2det}
\nu_{(N,\sigma)}(E_{\mathrm{ref}}) := 
\mathrm{sgn}\!\left( \det\!\left[ H_{(N,\sigma)} - E_{\mathrm{ref}}\1 \right] \right),
\end{align}
%%%%%%
with $E_{\mathrm{ref}} \in \mathbb{R}$. 

The above facts lead to a change of the zero-dimensional topological classification from $0$ 
to $\mathbb{Z}_2$ due to interactions, indicating the emergence of interaction-enabled EP2s [two-fold exceptional lines (EL2s)] in one- [two-] dimensional parameter space~\cite{Yoshida_PRB2019,Budich_PRB2019,Okugawa_PRB2019}.
In non-interacting cases, the nontrivial $\mathbb{Z}_2$ topology is forbidden [i.e., $\nu_{(N,\sigma)}(E_{\mathrm{ref}})>0$ for $E_{\mathrm{ref}}\in \mathbb{R}$] as follows from the relation $\det\!\left[ H_{(N,\sigma)} - E_{\mathrm{ref}}\1 \right]=\left|\det\!\left[H_{0+(N,\sigma)}-{E}_\mathrm{ref} \1 \right]\right|^2$ [see Eqs.~\eqref{eq: IEEL PT example} and \eqref{eq: IEEL free H example}].
On the other hand, in correlated cases, $\mathbb{Z}_2$ topology can be nontrivial because $H_{(N,\sigma)}$ is no longer block-diagonalized with respect to $\hat{T}_z$.
Here, we note that for systematic characterization~\cite{Wojcik_PRB2020,Yang_2021,Delplace_2021}, the following index
is more suited  because of its independence from the choice of $E_{\mathrm{ref}}$ 
%%%%%%
\begin{align}
\label{eq: IEEL Z2disc}
s_{(N,\sigma)}(\bm{\lambda}):= \mathrm{sgn}\!\Big(\mathrm{Disc}\big[
\mathrm{det}[H_{(N,\sigma)}(\boldsymbol{\lambda})-E\1]   \big]\Big),
\end{align}
%%%%%%
where ``$\mathrm{Disc}$" denotes the discriminant of a polynomial with respect to the variable $E$.
Both of $\mathbb{Z}_2$ indices $\nu_{(N,\sigma)}(E_{\mathrm{ref}})$ and $s_{(N,\sigma)}(\bm{\lambda})$ characterize the zero-dimensional topology of EP2s.
For details on the relation between Eqs.~\eqref{eq: IEEL Z2det} and
\eqref{eq: IEEL Z2disc}, see Sec.~\ref{sec: chara of PT} of Supplemental Material~\cite{Suppl}.

%%%%%%
\sectitle{
Interaction-enabled exceptional line
}
%%%%%%
By analyzing a bosonic toy model, we demonstrate the emergence of interaction-enabled EL2s in two-dimensional parameter space. The Hamiltonian reads
%%%%%%
\begin{align}
\label{eq: IEEL bosonic toy model}
&\hat{H}(x,y)=\hat{H}_{0\mathrm{A}}(x,y)+\hat{H}_{0\mathrm{B}}(x,y)+\hat{H}_{\mathrm{int}},\\
    &\hat{H}_{0\tau}=\hat{\Psi}^\dagger_{b,\tau} h_\tau(x,y)\hat{\Psi}_{b,\tau},\\
    &\hat{H}_{\mathrm{int}}=V\sum^2_{i=1}\big(\hat{b}^{\dagger}_{i,\mathrm{A}} \hat{b}_{i,\mathrm{B}}\hat{b}^{\dagger}_{3,\mathrm{A}} \hat{b}_{3,\mathrm{B}}+\mathrm{h.c.}\big)\nonumber\\
    & \ \ \ \ \ \ \ \ \  +U\big[(\hat{b}^{\dagger}_{1,\mathrm{A}} \hat{b}_{2,\mathrm{B}}+\hat{b}^{\dagger}_{2,\mathrm{A}} \hat{b}_{1,\mathrm{B}})\hat{b}^{\dagger}_{3,\mathrm{A}} \hat{b}_{3,\mathrm{B}}+\mathrm{h.c.}\big],
\end{align}
%%%%%%
with 
$\hat{\Psi}^\dagger_{b,\tau}=\big(\hat{b}^\dagger_{1,\tau},\ \hat{b}^\dagger_{2,\tau},\ \hat{b}^\dagger_{3,\tau}\big)$
and
$\hat{b}^\dagger_{i,\tau}$ ($\hat{b}_{i,\tau}$) 
creating (annihilating)
a boson with component $\tau \ (=\mathrm{A}, \mathrm{B})$ at  
site $i\ (=1,2,3)$.
The interaction strengths are denoted by 
real values $V$ and $U$.
The first-quantized Hamiltonian $h_\tau(x,y)$ is given by
%%%%%%
\begin{align}
    &h_\tau(x,y)=
    \begin{pmatrix}
        (-1)^{\delta_{\mathrm{B},\tau}}x-\ii\delta & y & 0\\
        y &  \ii\delta-(-1)^{\delta_{\mathrm{B},\tau}}x & 0\\
        0 & 0 & w
    \end{pmatrix},
\end{align}
%%%%%%
with the Kronecker delta 
$\delta_{\tau,\tau'}$ being equal to $1$ for $\tau=\tau'$ and $0$ otherwise.
The corresponding second-quantized Hamiltonian $\hat{H}(x,y)$ preserves charge U(1), pseudo-spin–parity, and $PT$ symmetries [Eqs.~\eqref{eq: IEEL c-U1 symm}–\eqref{eq: IEEL PT-H}].
The $PT$ operator is defined as
%%%%%%
\begin{align}
\hat{P}\hat{T}
&=
\exp\!\left[
\ii \frac{\pi}{2}
\sum_{i,j}
\left(
p_{ij}\hat{b}^\dagger_{i,\mathrm{A}}\hat{b}_{j,\mathrm{B}}
+
p^{T}_{ij}\hat{b}^\dagger_{i,\mathrm{B}}\hat{b}_{j,\mathrm{A}}
\right)
\right] \mathcal{K} ,
\end{align}
%%%%%%
with
$p=\begin{pmatrix}
            0 & 1 & 0\\
            1 & 0 & 0\\
            0 & 0 & 1
       \end{pmatrix}$.
In addition, both $\hat{H}(x,y)$ and $\hat{P}\hat{T}$ satisfy $[\hat{H}(x,y),\hat{n}_3]=0$ and $[\hat{P}\hat{T},\hat{n}_3]=0$
with $\hat{n}_3= \sum_{\tau=\mathrm{A},\mathrm{B}}\hat{b}^\dagger_{3,\tau}\hat{b}_{3,\tau}$.

%%%%%%%%%%%%%%%%%%%%%%%%%
\begin{figure}[!t]
\begin{minipage}{1\hsize}
\begin{center}
    \includegraphics[width=\linewidth,height=4.1cm,keepaspectratio]{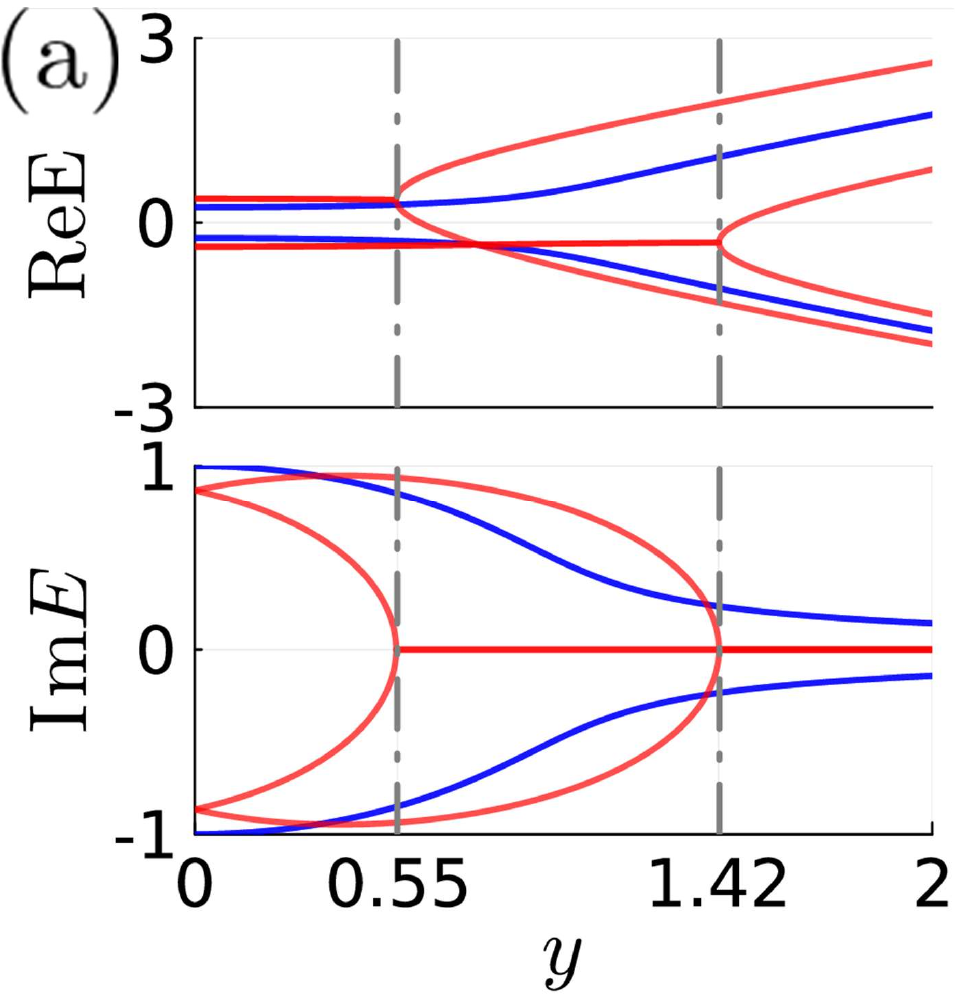}
    \includegraphics[width=\linewidth,height=4.1cm,keepaspectratio]{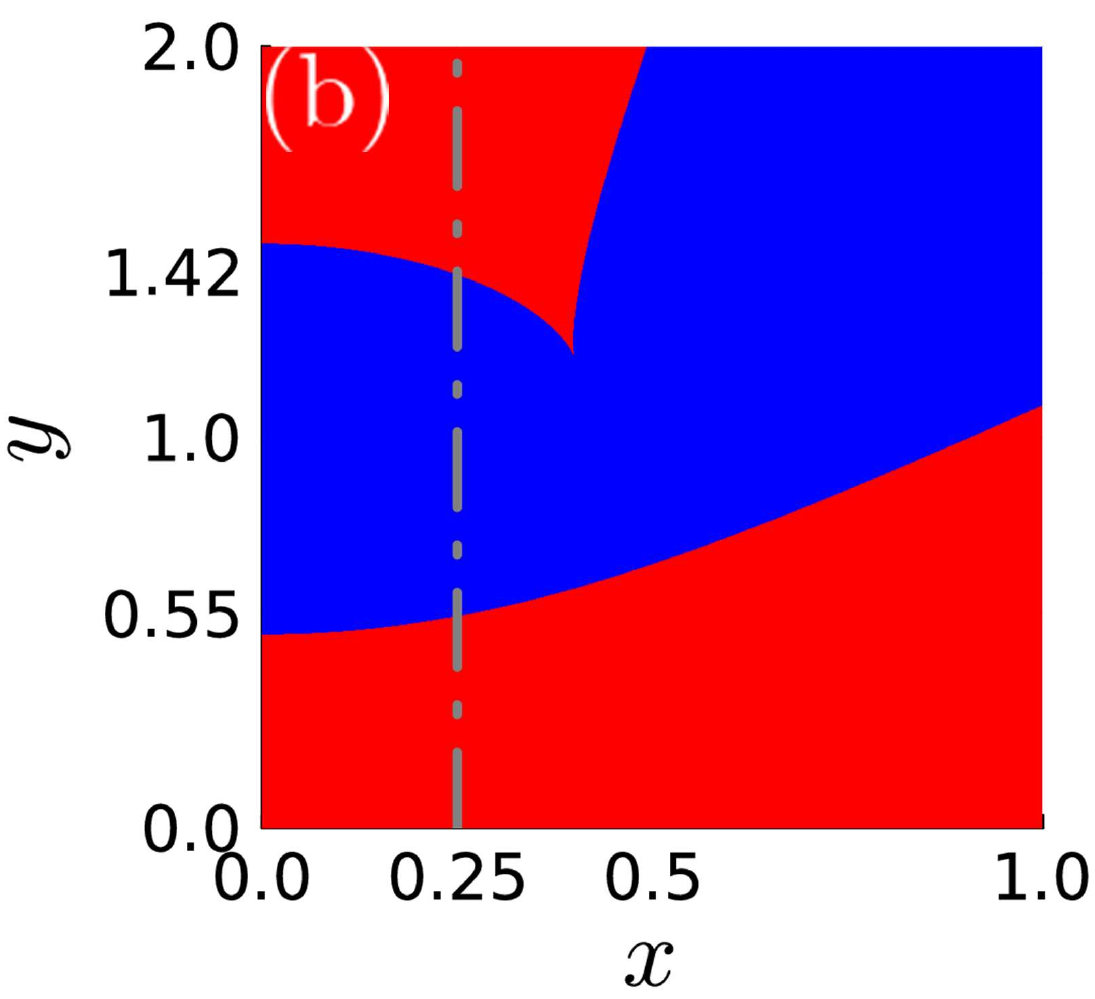}
\end{center}
\end{minipage}
\caption{
(a): $y$-dependence of the eigenvalues of $H_{(2,+)}(x=0.25,y)$ for $(w,\delta)=(0,1)$. The blue and red solid lines correspond to $V=U=0$ and $(V,U)=(0.3,0.5)$, respectively.
(b): The $\mathbb{Z}_2$ topological index $s_{(2,+)}(x,y)$ [see Eq.~\eqref{eq: IEEL Z2disc}] for $(w,\delta)=(0,1)$ and $(V,U)=(0.3,0.5)$. The blue and red regions correspond to $s_{(2,+)}=-1$ and $s_{(2,+)}=+1$, respectively.
}
\label{fig:IEEL boson Ey_s}
\end{figure}
%%%%%%%%%%%%%%%%%%%%%%%%%

We analyze the Hamiltonian acting within the Fock-space sector with $(N,\sigma)=(2,+)$ satisfying $N+1+\sigma = 0 \pmod{4}$ and containing one particle localized on site $i=3$.
The basis is chosen as
%%%%%%
\begin{align}
\label{eq: IEEP3 2-boson basis}
\Big(|\mathrm{A}1,\mathrm{A}3\rangle,|\mathrm{A}2,\mathrm{A}3\rangle,|\mathrm{B}1,\mathrm{B}3\rangle,|\mathrm{B}2,\mathrm{B}3\rangle\Big),
\end{align}
%%%%%%
with states defined as 
%%%%%%
\begin{align}
|\tau i,\tau 3\rangle=\hat{b}^\dagger_{i,\tau}\hat{b}^\dagger_{3,\tau}|0\rangle.
\end{align}
%%%%%%
In this Fock-space sector, the Hamiltonian is represented as 
%%%%%%
\begin{align}
\label{eq: IEEP3 2-boson hamil}
H_{(2,+)}=\begin{pmatrix}
        H_{0(2,+,1)} & O \\O & H_{0(2,+,-1)}
    \end{pmatrix}+H_{\mathrm{int}(2,+)}.
\end{align}
%%%%%%
%
Here, $H_{0(2,+,T_z)}$ ($T_z=\pm1$) is given by
%%%%%%
\begin{align}
   H_{0(2,+,T_z)}=w\1_2+y\sigma_1+\left[(-1)^{\delta_{T_z,-1}}x-\ii\delta\right]\sigma_3,
\end{align}
%%%%%%
with $\sigma_1$ and $\sigma_3$ are the Pauli matrices.
The interaction term $H_{\mathrm{int}(2,+)}$ is given by
%%%%%%
\begin{align}
H_{\mathrm{int}(2,+)}=
V
\begin{pmatrix}
  &  & 1& \\
  &  &  & 1 \\
1 &  & & \\
  & 1& & \\
\end{pmatrix}
+
U
\begin{pmatrix}
 & & & 1\\
 & & 1& \\
 & 1& & \\
1 & & & \\
\end{pmatrix}.
\end{align}
%%%%%%
Zero matrix elements are omitted for brevity.

As an illustrative example 
\footnote{
Since the spectrum is symmetric about $x$ and $y$ axes, we focus on the first quadrant\hspace{-3pt}
}, 
we choose $x=0.25$ and plot the $y$-dependence of the complex eigenvalues of $H_{(2,+)}(x=0.25,y)$ in Fig.~\ref{fig:IEEL boson Ey_s}(a). For $V=U=0$, no degeneracy of the complex eigenvalues occurs. 
In contrast, for $(V,U)=(0.3,0.5)$, 
degeneracies are observed at $y\sim 0.55$ and $y\sim 1.42$, indicating the emergence of interaction-enabled EP2s.
At these points, the number of complex-conjugate eigenvalue pairs changes; as $y$ increases, the number of pairs changes from two to one at $y\sim 0.55$ and from one to zero at $y\sim 1.42$ [see Fig. 2(a)]. Correspondingly, the topological index $s_{(2,+)}(x=0.25,y)$ 
changes from $+1$ to $-1$ at $y\sim 0.55$ and from $-1$ to $1$ at $y\sim 1.42$ [see Fig.~\ref{fig:IEEL boson Ey_s}(b)], elucidating the topology of interaction-enabled EP2s.
In addition, Fig.~\ref{fig:IEEL boson Ey_s}(b) indicates that 
these points extend to other values of $x$, forming the interaction-enabled EL2s.

The above results demonstrate the emergence of interaction-enabled EL2s characterized by a topological $\mathbb{Z}_2$ index in a two-component bosonic system possessing charge U(1), pseudo-spin–parity, and $PT$ symmetries.

%%%%%%
\sectitle{Dynamical properties}
%%%%%%
%%%%%%%%%%%%%%%%%%%%%%%%%
\begin{figure}[!b]
\begin{minipage}{1\hsize}
\begin{center}
    \includegraphics[width=0.49\linewidth,height=4.1cm,keepaspectratio]{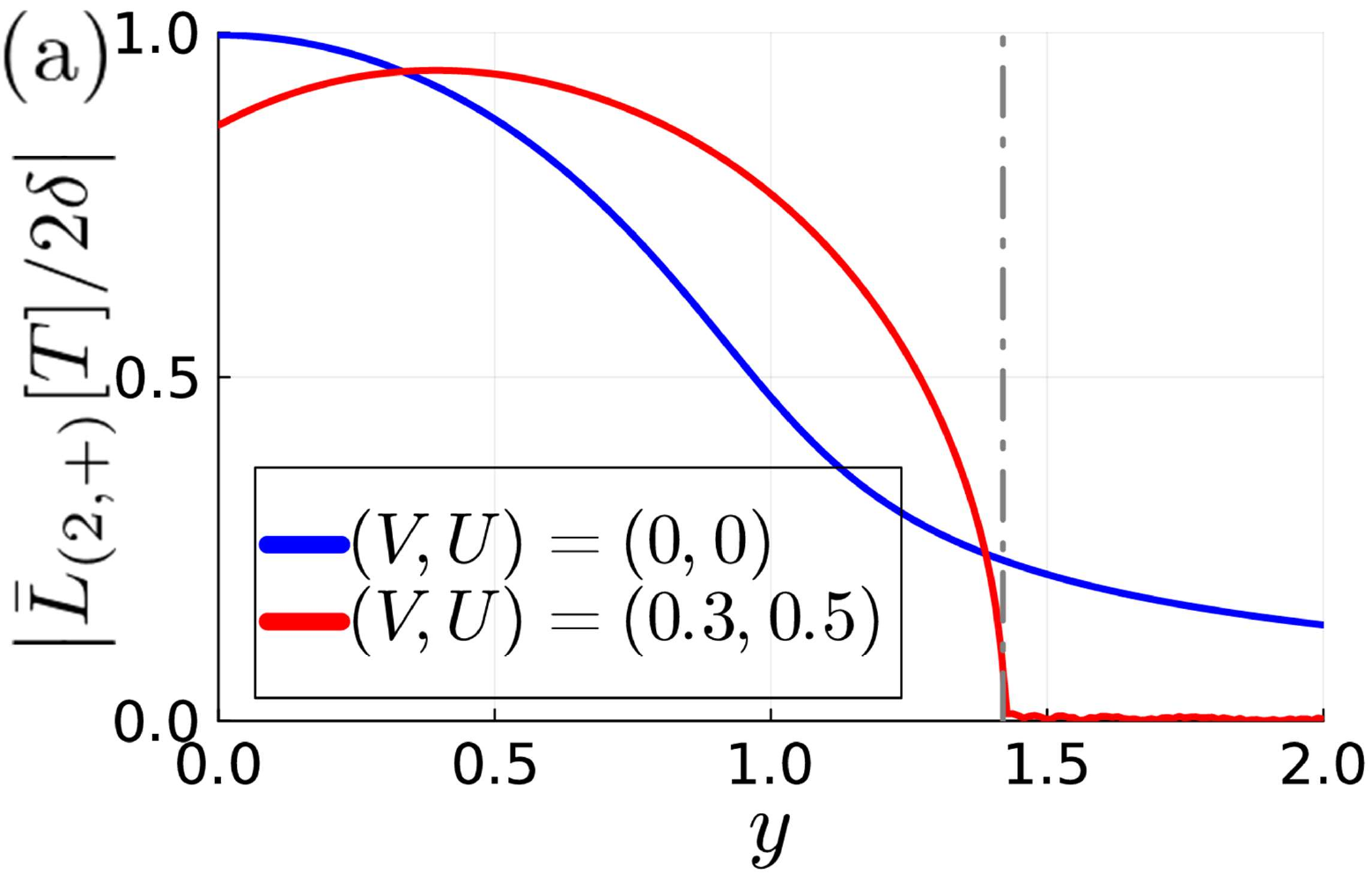}
    \includegraphics[width=0.49\linewidth,height=4.1cm,keepaspectratio]{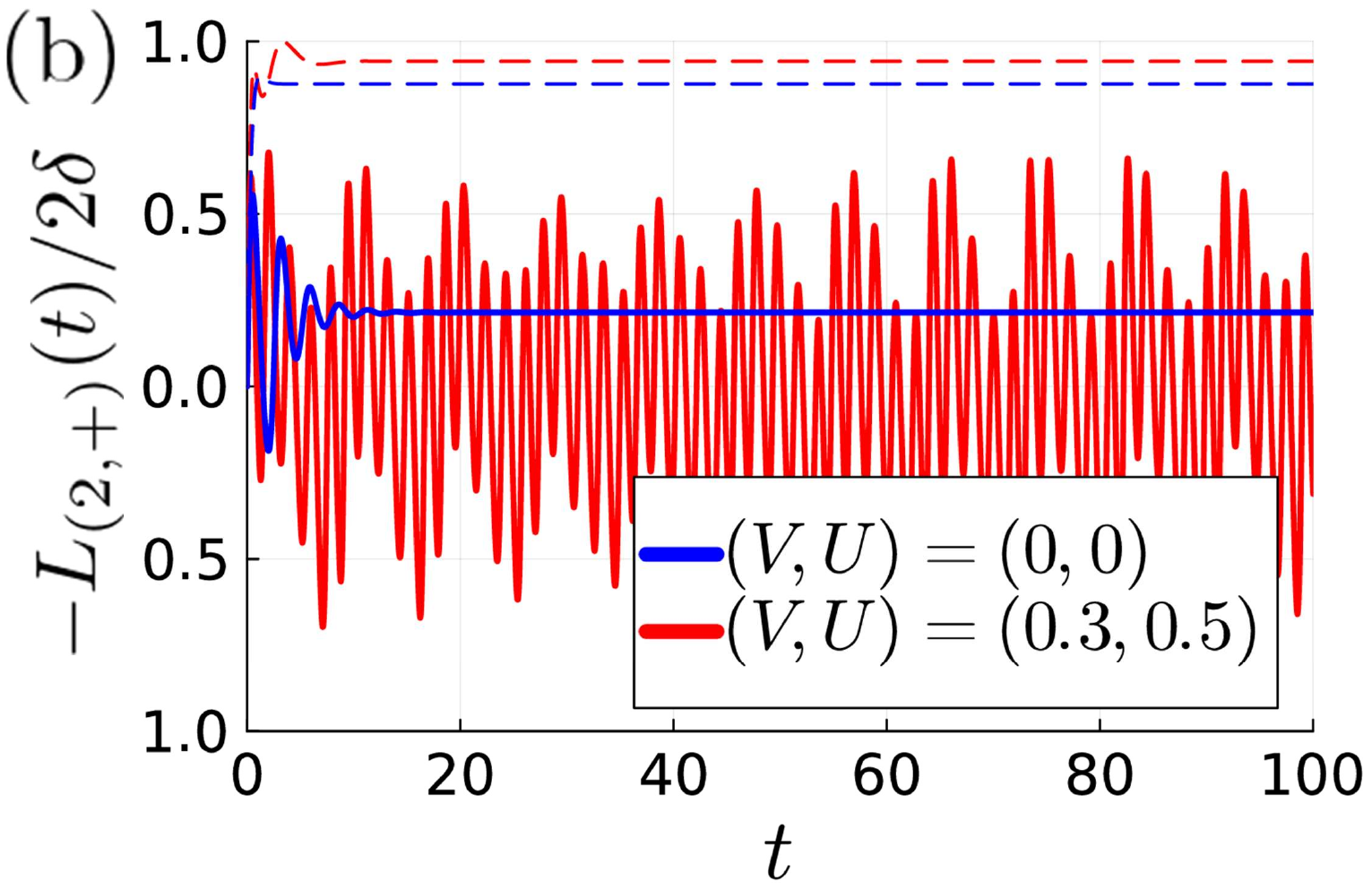}
\end{center}
\end{minipage}
\caption{
(a): Dependence of $|\bar{L}_{(2,+)}[T]/2\delta|$ on $y$ for $(w,\delta,x,T)=(0,1,0.25,100)$. The gray dashed line indicates $y=1.42$.
(b): Time dependence of $-{L}_{(2,+)}(t)/2\delta$ for $(w,\delta,x)=(0,1,0.25)$. The dashed and solid lines correspond to $y=0.5$ and $y=1.5$, respectively. In panels (a) and (b), blue and red denote the cases $V=U=0$ and $(V,U)=(0.3,0.5)$, respectively.
}
\label{fig:IEEL_b_mean_L_Lt}
\end{figure}
%%%%%%%%%%%%%%%%%%%%%%%%%
The existence of interaction-enabled EL2s protected by $PT$ symmetry 
is reflected in the behaviors of loss rate, an experimentally measurable quantity for cold atoms~\cite{Takasu_ColdAtom_PTEP2020}.
In the Fock-space sector labeled by $\left(N,(-1)^{N_\mathrm{B}}\right)=(2M,\sigma)$,
the loss rate of a time-independent Hamiltonian $\hat{H}(\boldsymbol{\lambda})$ is defined as follows:
%%%%%%
\begin{align}
    \label{eq: IEEL loss rate}
    \langle{N_{(2M,\sigma)}(t)}\rangle&:=\mathrm{Tr}[\hat{N}\mathcal{P}_{(2M,\sigma)}\hat{\rho}(t)\mathcal{P}_{(2M,\sigma)}],\\
    L_{(2M,\sigma)}(t)&:=-\frac{1}{\langle N_{(2M,\sigma)}(t)\rangle}\frac{d}{dt}\langle{N_{(2M,\sigma)}(t)}\rangle,
\end{align}
%%%%%%
where we define the density matrix as $\hat{\rho}(t):= e^{-\ii t\hat{H}}|\psi(0)\rangle\langle\psi(0)|e^{\ii t\hat{H}^\dagger}$, and $\mathcal{P}_{(2M,\sigma)}$ is the projection operator onto the Fock-space sector with $(2M,\sigma)$.
We also introduce the time-averaged loss rate from $t=0$ to $t=T$ as
%%%%%%
\begin{align}
    \label{eq: IEEL loss rate ave}
    \bar{L}_{(2M,\sigma)}[T]:=\frac{1}{T}\int^T_0dt \ L_{(2M,\sigma)}(t).
\end{align}
%%%%%%
We set the initial state as $|\psi(0)\rangle=\frac{1}{\sqrt{2}}\big(|\mathrm{A}2,\mathrm{A}3\rangle+\ii|\mathrm{B}1,\mathrm{B}3\rangle\big)$
and discuss the behavior of the loss rate at $x=0.25$, the same value as in Fig.~\ref{fig:IEEL boson Ey_s}(a).

Figure~\ref{fig:IEEL_b_mean_L_Lt}(a)
shows that, in the interacting case, $\bar{L}_{(2,+)}[T]$ rapidly approaches zero near $y\sim 1.42$, where an interaction-enabled EP2 emerges, and remains approximately zero for $y\gtrsim 1.42$.
This behavior is due to the fact that the time-reversal loss rate satisfies $-\bar{L}_{(2,+)}[T]\sim 2\gamma_{(2,+)}$ for sufficiently large $T$ with $\gamma_{(2,+)}$ being the maximum of the imaginary parts of the complex eigenvalues of $H_{(2,+)}$ (for further details on the loss rate, see Sec.~\ref{sec: chara of PT} of Supplemental Material~\cite{Suppl}).
These data indicate that the emergence of the interaction-enabled EL2 is reflected in the loss rate. This fact can also be found in the time-dependence of the loss rate [see Fig.~\ref{fig:IEEL_b_mean_L_Lt}(b)].

%%%%%%
\sectitle{Interaction-enabled EP3}
%%%%%%
In the above, we have discussed the emergence of interaction-enabled EL2s in two-dimensional parameter space, in which interactions play an essential role.
We further extend this argument and reveal the emergence of interaction-enabled EP3s in two-dimensional parameter space, which is protected by topology 
beyond the point-gap topological classifications.

\begin{figure}[!t]
\begin{minipage}{0.55\hsize}
\begin{center}
\includegraphics[width=\linewidth,height=6cm,keepaspectratio]{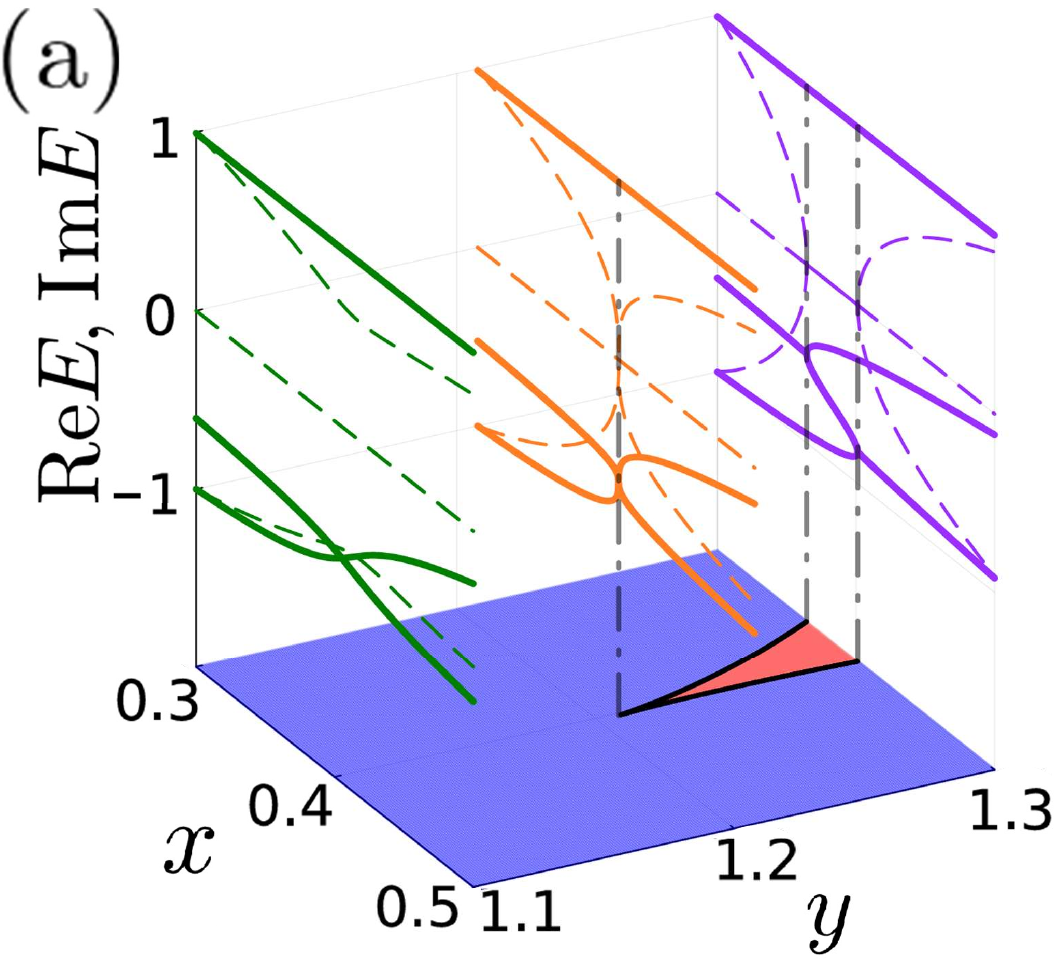}
\end{center}
\end{minipage}
\begin{minipage}{0.43\hsize}
\begin{center}    \includegraphics[width=\linewidth,height=3.5cm,keepaspectratio]{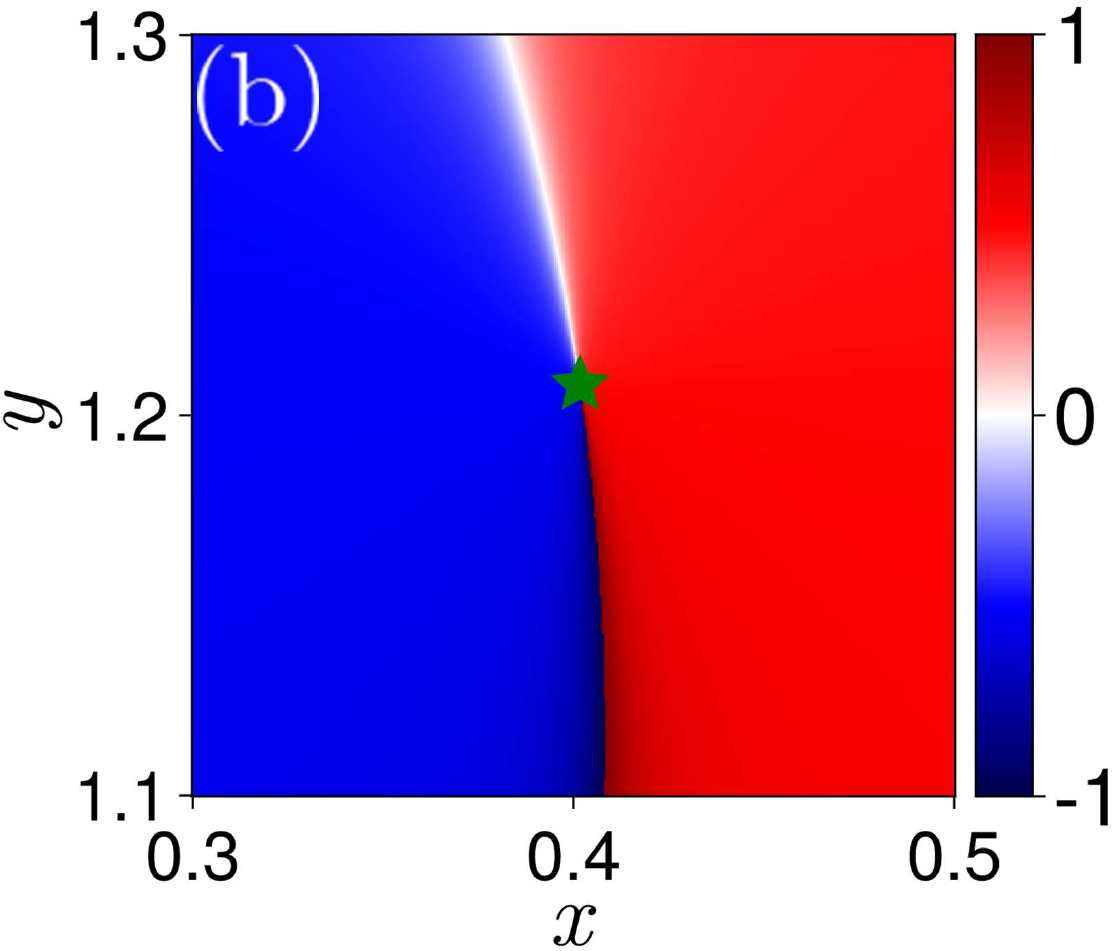}
\end{center}
\end{minipage}
\caption{
(a): Real (solid) and imaginary (dashed) parts of the complex eigenvalues of $H_{(2,+)}$ as functions of $x$ for $y=1.1$, $1.208$, and $1.3$.
For each value of $y$, the spectra are normalized to $\pm1$ over $x\in[0.3,0.5]$.
Blue and red regions on the bottom plane indicate $s_{(2,+)}=-1$ and $+1$, respectively [see Fig.~\ref{fig:IEEL boson Ey_s}]; black lines denote the phase boundaries, whose intersection gives $\boldsymbol{\lambda}_{\mathrm{EP3}}\sim(0.402,1.208)$.
(b): Principal value of $\mathrm{arg} [Z_r]/\pi$ in the $x$–$y$ plane.
The green star marks $\boldsymbol{\lambda}_{\mathrm{EP3}}$.
These data are obtained for $(w,\delta,V,U)=(0,1,0.3,0.5)$.
}
\label{fig:IEEP3}
\end{figure}

We consider two distinct interaction-enabled EL2s in two-dimensional parameter space, each of which corresponds to the degeneracy of a different pair of eigenvalues. Specifically, we assume that eigenvalues $E_1$ and $E_2$ are degenerate along one EL2, while $E_2$ and $E_3$ are degenerate along the other. At the point where these two EL2s intersect, three eigenvalues $E_1$, $E_2$, and $E_3$ become degenerate, resulting in an interaction-enabled EP$3$.
Indeed, such a situation is observed for our toy model [see Fig.~\ref{fig:IEEP3}(a)].
We stress that in the Fock-space sector with $(N,\sigma)$ satisfying $N+1+\sigma = 0 \pmod{4}$, EP3s are  
interaction-enabled,
because they emerge as intersections of interaction-enabled EL2s, which is consistent with the codimension argument.

%%%%%%%%%%%%%%%%%%%%%%%%%
%%%%%%%%%%%%%%%%%%%%%%%%%

The topology of the interaction-enabled EP3 is characterized by the resultant winding number.
Let us consider the characteristic polynomial $P(E;\boldsymbol{\lambda})=\det[H(\boldsymbol{\lambda})-E\1_n]$ of a $PT$-symmetric non-Hermitian matrix $H(\boldsymbol{\lambda}) \in \mathbb{C}^{n\times n} \ (n\geq 3)$.
If $H(\boldsymbol{\lambda})$ possesses an EP3 at $\boldsymbol{\lambda}=\boldsymbol{\lambda}_{\mathrm{EP3}}$, and the corresponding eigenvalue is $E_{\mathrm{EP3}} \in \mathbb{R}$, then the characteristic polynomial has a triple root at $E=E_{\mathrm{EP3}}$. 
The topology of the triple root can be characterized by the resultant winding number 
%%%%%%
\begin{align}
\label{eq: IEEP3 inv}
    W_r&=\oint\frac{d\boldsymbol{\lambda}}{2\pi \ii}\cdot\partial_{\boldsymbol{\lambda}} \ \mathrm{ln} \ Z_r(\boldsymbol{\lambda}),
\end{align}
%%%%%%
where $Z_r$ is defined as
$Z_r(\boldsymbol{\lambda})=r_1(\boldsymbol{\lambda})+\ii r_2(\boldsymbol{\lambda})$ with
%%%%%%
\begin{align}
    \label{eq: EP3 resultant com}
    r_j(\boldsymbol{\lambda})&:=\mathrm{Res}\!\left[\partial^{2-j}_{\tilde{E}} P_\mathrm{EP3}(\tilde{E};\boldsymbol{\lambda}),\partial^{2}_{\tilde{E}} P_\mathrm{EP3}(\tilde{E};\boldsymbol{\lambda})\right],
\end{align}
%%%%%%
for $j=1,2$. 
The integral is taken over a closed path enclosing $\bm{\lambda}=\bm{\lambda}_{\mathrm{EP3}}$.
Here, $\mathrm{Res}[f,g]$ denotes the resultant of two polynomials $f$ and $g$. Due to $PT$ symmetry, $r_j$ ($j=1,2$) are real.
The polynomial of degree three $P_\mathrm{EP3}(\tilde{E};\boldsymbol{\lambda})$ with $\tilde{E}=E-E_{\mathrm{EP3}}$ are obtained by the Taylor expansion of $P(E;\boldsymbol{\lambda})$
around $\tilde{E}=0$:
%%%%%%
\begin{align}
 P_\mathrm{EP3}(\tilde{E};\boldsymbol{\lambda})
    :=\sum^3_{k=0}e'_{n-k}(\boldsymbol{\lambda})(-1)^k\tilde{E}^k
\end{align}
%%%%%%
with $e'_{k+1}$ recursively obtained from 
%%%%%%
\begin{align*}
    (k+1)e'_{k+1}(\boldsymbol{\lambda})&=\sum^k_{l=0}(-1)^l \ e'_{k-l}(\boldsymbol{\lambda}) \ \Tr\left[\big(H'(\boldsymbol{\lambda})\big)^{l+1}\right],
\end{align*}
%%%%%%
and $H'(\boldsymbol{\lambda})=H(\boldsymbol{\lambda})-E_{\mathrm{EP3}}\1_n$ (for more details see Sec.~\ref{sec: Newtons I} of Supplemental Material~\cite{Suppl}).
The non-zero resultant winding number indicates that the resultants vanish $(r_1,r_2)=\bm{0}$ at a point inside the closed loop where $P_\mathrm{EP3}(\tilde{E};\boldsymbol{\lambda})$ has a triple root~\cite{Delplace_2021}. 
Computing the resultant winding number, we obtain $W_{\mathrm{r}}=-1$ [see Fig.~\ref{fig:IEEP3}(b)], elucidating the topological protection of the interaction-enabled EP3. 

We finish this part with three remarks.
(i) We note that our argument suggests a border class of interaction-enabled EP$n$s.
For instance, it is straightforward to generalize interaction-enabled EP4s in three-dimensional parameter space.
(ii) We also note that while the resultant winding number requires the Taylor expansion around $E_{\mathrm{EP3}}$, alternative winding number enables the characterization of EP$n$s for an $(n+1)\times (n+1)$ matrix without such an expansion for $n\geq 3$, thereby allowing a systematic analysis (for more details see Secs.~\ref{sec: chara multi-fold EP} and~\ref{sec: nume data for boson} of Supplemental Material~\cite{Suppl}).
(iii) In the above, although we have restricted ourselves to bosonic systems, a similar analysis demonstrates the emergence of interaction-enabled topology in fermionic systems (for more details, see Sec.~\ref{sec: results of fermionic sys} of Supplemental Material~\cite{Suppl}).

%%%%%%
\sectitle{Summary}
%%%%%%
%
We have proposed a novel type of non-Hermitian degeneracies,  
dubbed interaction-enabled EP$n$s ($n=2,3$) ---EP$n$s protected by topology that are
prohibited at the non-interacting level. Specifically, analyzing second-quantized Hamiltonians, we have demonstrated that both bosonic and fermionic 
systems can host such interaction-enabled EP$n$s in parameter space that are protected by charge U(1), pseudo-spin-parity, and $PT$ symmetries. Interaction-enabled EP2s are protected 
by zero-dimensional topology and give rise to qualitative changes in the loss rate, an experimentally measurable quantity for 
cold atoms.
Furthermore, we have revealed that interactions enable EP3s 
protected by topology beyond the point-gap topological classifications, which suggest the potential presence of a broader class of interaction-enabled EP$n$s. 

We conclude this paper by noting that identifying feasible platforms for interaction-enabled EP$n$s is an important step toward their experimental observation. Among the various non-Hermitian systems, cold atoms are considered as a potential platform where an EP3 in parameter space is reported~\cite{Wang_PRL2024}. 
A detailed analysis along this direction is left for future work.

%%%%%%
\sectitle{Acknowledgments}
%%%%%%
This work is supported by JSPS KAKENHI Grant Nos.~JP21K13850, JP23KK0247, JP25K07152, and JP25H02136, as well as JSPS Bilateral Program No.~JPJSBP120249925.
T.~Y.~is grateful for the support from the ETH Pauli Center for Theoretical Studies and the Grant from Yamada Science Foundation.

%%%%%%%%%%%%%%%
\input{./main.bbl}
%%%%%%%%%%%%%%%
%

\clearpage
%%%%%%%%%%%%%%%%%%%%%%%%%%%%%
%%% SUPPLEMENTAL MATERIAL %%%
%%%%%%%%%%%%%%%%%%%%%%%%%%%%%

\renewcommand{\thesection}{S\arabic{section}}
\renewcommand{\theequation}{S\arabic{equation}}
\setcounter{equation}{0}
\renewcommand{\thefigure}{S\arabic{figure}}
\setcounter{figure}{0}
\renewcommand{\thetable}{S\arabic{table}}
\setcounter{table}{0}
\makeatletter
\c@secnumdepth = 2
\makeatother

\onecolumngrid
\thispagestyle{empty}
\begin{center}
 {\large \textmd{Supplemental Material:} \\[0.3em]
 {\bfseries 
  Interaction-Enabled Two- and Three-Fold Exceptional Points
 }
 }
\end{center}

\setcounter{page}{1}
\setcounter{section}{0}

%%%%%%
\section{Symmetries and topology on the Fock-space sector with $(N,\sigma)$} \label{sec: Fock-sp}
%%%%%%
Here, we consider both bosonic and fermionic non-Hermitian systems possessing charge U(1), pseudo-spin-parity, and $PT$ symmetries.
Accordingly, the second-quantized Hamiltonian $\hat{H}=\hat{H}_0+\hat{H}_{\mathrm{int}}$ satisfies the following relations:
%%%%%%
\begin{align}
\label{eq: app IEEL c-U1 symm}
&[\hat{H},\hat{N}]=0 ,\\
\label{eq: app IEEL spin-parity symm}
&[\hat{H},(-1)^{\hat{T}_z}]=0 ,\\
\label{eq: app IEEL PT symm}
&[\hat{P}\hat{T},\hat{H}]=0 .
\end{align}
%%%%%%
Here, we assume that the $PT$ operator satisfies
%%%%%%
\begin{align}
(\hat{P}\hat{T})^2&=\1 , \\
\label{eq: PT-N}
[\hat{P}\hat{T},\hat{N}]&=0 ,\\
\label{eq: PT-Tz}
\{\hat{P}\hat{T},\hat{T}_z\}&=0 .
\end{align}
%%%%%%
The total particle-number operator $\hat{N}$ and the pseudo-spin operator $\hat{T}_z$ are defined as
%%%%%%
\begin{align}
\hat{N}&:= \hat{N}_{\mathrm{A}}+\hat{N}_{\mathrm{B}},\\
\hat{T}_z&:= \frac{1}{2}(\hat{N}_{\mathrm{A}}-\hat{N}_{\mathrm{B}}),\\
\hat{N}_\tau&:= \sum_i \hat{a}^\dagger_{i,\tau}\hat{a}_{i,\tau}.
\end{align}
%%%%%%
The operators $\hat{a}^\dagger_{i,\tau}$ and $\hat{a}_{i,\tau}$ denote the creation and annihilation operators of particles in component $\tau$ $(=\mathrm{A},\mathrm{B})$, respectively.
These operators obey bosonic commutation relations or fermionic anticommutation relations, depending on the statistics of the particles.

Under the above symmetry conditions, $\hat{H}$ commutes with $\hat{N}$ and $(-1)^{\hat{N}_{\mathrm{B}}}\ (\hat{N}_\mathrm{B}=\hat{N}/2-\hat{T}_z)$, and can hence be block-diagonalized as
%%%%%%
\begin{align}
\hat{H}=\bigoplus_{N=0,1,\dots}\bigoplus_{\sigma=\pm} H_{(N,\sigma)},
\end{align}
%%%%%%
where $N$ and $\sigma$ denote the eigenvalues of $\hat{N}$ and $(-1)^{\hat{N}_{\mathrm{B}}}$, respectively.
Furthermore, in the absence of interactions, $\hat{H}=\hat{H}_0$ can be expressed as a quadratic form of the creation and annihilation operators.  
From the commutation of $\hat{H}$ with $(-1)^{\hat{T}_z}$, it follows that the corresponding first-quantized free-particle Hamiltonian $h$ commutes with the first-quantized pseudo-spin operator $t_z$, i.e.,
$[h,t_z]=0$.  
This implies $[\hat{H},\hat{T}_z]=0$, which in turn allows $H_{(N,\sigma)}$ to be further block-diagonalized as
%%%%%%
\begin{align}
\label{eq: non-int add symm app}
H_{(N,\sigma)}=\bigoplus_{T_z} H_{0(N,\sigma,T_z)},
\end{align}
%%%%%%
where $T_z$ denotes the eigenvalues of $\hat{T}_z$.

Considering the zero-dimensional point-gap topology of the Hamiltonian $H_{(N,\sigma)}$, we find
%%%%%%
\begin{align}
\label{eq: change of 0-dim topo}
& 0 \rightarrow\mathbb{Z}_2 \quad\quad \mathrm{for}\ N+1+\sigma=0 \pmod{4}, \\
& \mathbb{Z}_2 \rightarrow\mathbb{Z}_2 \quad\quad \mathrm{for}\ N+1+\sigma=2 \pmod{4}, \\
& 0 \rightarrow0 \quad\quad \mathrm{for}\ N+1+\sigma=1,3 \pmod{4},
\end{align}
%%%%%%
where the topology on the left (right) side of each arrow corresponds to the non-interacting (interacting) case.
In the following, we explain the details of this result.

%%%%%%
\subsection{$PT$ symmetry and the Fock-space sector labeled by $(N,\sigma)$}
\label{subsec: PT in N_sigma-space}
%%%%%%
Firstly, we analyze the action of $\hat{P}\hat{T}$ on the subspace labeled by $(N,\sigma)$. 
From Eqs.~\eqref{eq: PT-N} and \eqref{eq: PT-Tz}, we obtain
%%%%%%
\begin{align}
\label{eq: IEEL H-PT algebraic relation}
\hat{P}\hat{T}(-1)^{\hat{N}_\mathrm{B}}(\hat{P}\hat{T})^{-1} = (-1)^{\hat{N}}(-1)^{\hat{N}_\mathrm{B}}.
\end{align}
%%%%%%
It then follows that the action of $\hat{P}\hat{T}$ is closed within the $(N,\sigma)$-labeled Fock-space sector when $N$ is even, while it is not closed when $N$ is odd.

Let us consider the case where $N$ is even. 
In this case, the action of $\hat{P}\hat{T}$ is closed within the Fock-space sector labeled by $(N,\sigma)$. 
However, for the Fock-space sector labeled by $(N,\sigma,T_z)$, the anticommutation relation $\{\hat{P}\hat{T}, \hat{T}_z\}=0$ implies that the action of $\hat{P}\hat{T}$ maps the $(N,\sigma,T_z)$ subspace to the $(N,\sigma,-T_z)$ subspace. 
In other words, only the Fock-space sector with finite $T_z=0$ is closed under the $PT$ operation.  
Moreover, due to $PT$ symmetry, the bases of the $(N,\sigma,T_z)$ and $(N,\sigma,-T_z)$ subspaces are in one-to-one correspondence, and thus the dimensions of these subspaces are equal.
Next, we examine under which conditions the $(N,\sigma)$-labeled Fock-space sector contains a $T_z=0$ subspace. 
By introducing a non-negative integer $M$ such that $N=2M$, $T_z$ can be expressed as
$T_z = M - N_\mathrm{B}$,
 where $N_\mathrm{B}$ denotes the eigenvalues of $\hat{N}_\mathrm{B}$. 
In a subspace with fixed $(-1)^{N_\mathrm{B}}=\sigma$, $N_\mathrm{B}$ can take either only even or only odd values.
From this analysis, the $(N,\sigma)$-labeled Fock-space sector does not contain a $T_z=0$ subspace in the following cases: 
(i) $M$ is even and $\sigma=-$, or 
(ii) $M$ is odd and $\sigma=+$. 
These cases can be collectively expressed as $N+1+\sigma = 0\pmod{4}$.  
On the other hand, the $(N,\sigma)$-labeled Fock-space sector contains a $T_z=0$ subspace in the following cases: 
(i) $M$ is even and $\sigma=+$, or 
(ii) $M$ is odd and $\sigma=-$, 
which can be collectively expressed as $N+1+\sigma = 2\pmod{4}$.

For odd $N$, as stated at the beginning of this section, the action of $\hat{P}\hat{T}$ is not closed within the $(N,\sigma)$ Fock-space sector. 
In this case, we have $N+1+\sigma = 1,3 \pmod{4}$.

We next provide an explicit example of a many-body $\hat{P}\hat{T}$ operator satisfying $(\hat{P}\hat{T})^2=\1$, $[\hat{P}\hat{T},\hat{N}]=0$ and $\{\hat{P}\hat{T},\hat{T}_z\}=0$.  
Let $p$ be a real orthogonal matrix ($p^T p = \1$), and consider
%%%%%%
\begin{align}
\label{eq: PT define} 
\hat{P}\hat{T} &= e^{\ii \frac{\pi}{2} \hat{Q}} \mathcal{K}, \\
\hat{Q} &= \sum_{i,j} \big( p_{ij} \hat{a}^\dagger_{i,\mathrm{A}} \hat{a}_{j,\mathrm{B}} + p^T_{ij} \hat{a}^\dagger_{i,\mathrm{B}} \hat{a}_{j,\mathrm{A}} \big),
\end{align}
%%%%%%
where $\mathcal{K}$ denotes the complex conjugation operator, and $\hat{a}^\dagger_{j,\tau}$ ($\hat{a}_{j,\tau}$) denotes the creation (annihilation) operator for a boson or fermion in component $\tau$ at site $j$.
Since $\hat{Q}$ is real, $\hat{Q} = \hat{Q}^*$, it follows that
%%%%%%
\begin{align}
(\hat{P}\hat{T})^2 = e^{\ii \frac{\pi}{2} \hat{Q}} e^{-\ii \frac{\pi}{2} \hat{Q}}\mathcal{K}^2 = \1.
\end{align}
%%%%%%
Furthermore, one finds
%%%%%%
\begin{align}
\hat{P}\hat{T} \hat{a}^\dagger_{i,\mathrm{A}} (\hat{P}\hat{T})^{-1} 
&= e^{\ii \frac{\pi}{2} \hat{Q}} \mathcal{K} \hat{a}^\dagger_{i,\mathrm{A}} \mathcal{K} e^{-\ii \frac{\pi}{2} \hat{Q}} \nonumber\\
&= \hat{a}^\dagger_{i,\mathrm{A}} + \frac{\ii \pi}{2} [\hat{Q}, \hat{a}^\dagger_{i,\mathrm{A}}] 
   + \frac{(\frac{\ii \pi}{2})^2}{2!} \big[\hat{Q}, [\hat{Q}, \hat{a}^\dagger_{i,\mathrm{A}}]\big] + \cdots \nonumber\\
&= \cos\Big(\frac{\pi}{2}\Big) \hat{a}^\dagger_{i,\mathrm{A}} 
   + \ii \sin\Big(\frac{\pi}{2}\Big) \sum_j \hat{a}^\dagger_{j,\mathrm{B}} p^T_{ji} \nonumber\\
&= \ii \sum_j \hat{a}^\dagger_{j,\mathrm{B}} p^T_{ji}.
%%%%%%
\end{align}
Similarly, one obtains
%%%%%%
\begin{align}
\hat{P}\hat{T}
\begin{pmatrix}
\hat{a}^\dagger_{i,\mathrm{A}} \\ \hat{a}^\dagger_{i,\mathrm{B}}
\end{pmatrix}
(\hat{P}\hat{T})^{-1}
&= \ii \sum_j
\begin{pmatrix}
\hat{a}^\dagger_{j,\mathrm{B}} p^T_{ji} \\
\hat{a}^\dagger_{j,\mathrm{A}} p_{ji}
\end{pmatrix}, \\
\hat{P}\hat{T}
\begin{pmatrix}
\hat{a}_{i,\mathrm{A}} \\ \hat{a}_{i,\mathrm{B}}
\end{pmatrix}
(\hat{P}\hat{T})^{-1}
&= - \ii \sum_j
\begin{pmatrix}
p_{ij} \hat{a}_{j,\mathrm{B}} \\
p^T_{ij} \hat{a}_{j,\mathrm{A}}
\end{pmatrix}.
\end{align}
%%%%%%
From these expressions, it is straightforward to verify that the $\hat{P}\hat{T}$ operator defined in Eq.~\eqref{eq: PT define} satisfies $[\hat{P}\hat{T}, \hat{N}] = 0$ and $\{\hat{P}\hat{T}, \hat{T}_z\} = 0$.

%%%%%%
\subsection{Zero-dimensional point-gap topology of $H_{(N,\sigma)}$}
%%%%%%
We now discuss the zero-dimensional point-gap topology of $H_{(N,\sigma)}$, 
which acts invariantly within each Fock-space sector labeled by $(N,\sigma)$.
The associated topological index is defined as
%%%%%%
\begin{align}
\label{eq: Z2inv app}
\nu_{(N,\sigma)} 
= \mathrm{sgn}\!\left(
\det\!\left[H_{(N,\sigma)} - E_{\mathrm{ref}} \1 \right]
\right),
\end{align}
%%%%%%
where $E_{\mathrm{ref}}$ is a real reference energy.

In the absence of many-body interactions, the Hamiltonian $\hat{H}=\hat{H}_0$,
in addition to satisfying the symmetry conditions
Eqs.~\eqref{eq: app IEEL c-U1 symm}-\eqref{eq: app IEEL PT symm},
also obeys
%%%%%%
\begin{align}
[\hat{H}_0, \hat{T}_z] = 0 .
\end{align}
%%%%%%
As a consequence, $H_{(N,\sigma)}$ can be further block-diagonalized as in
Eq.~\eqref{eq: non-int add symm app}.
As we shall see below, this additional symmetry satisfied by 
$\hat{H}_0$ leads to a qualitative difference in the point-gap topology
between the noninteracting and interacting cases.
As discussed in Sec.~\ref{subsec: PT in N_sigma-space},
the constraints imposed by $\hat{P}\hat{T}$ on the Hamiltonian
within each Fock-space sector labeled by $(N,\sigma)$
depend on the value of $N+1+\sigma \ (\mathrm{mod}\,4)$.
We therefore analyze the point-gap topology separately for each case.

For the case $N+1+\sigma = 0 \pmod{4}$, the action of $\hat{P}\hat{T}$ is closed within the Fock-space sector labeled by $(N,\sigma)$.
In the absence of interactions, the Hamiltonian can be further block-diagonalized by $\hat{T}_z$, but there exists no subspace corresponding to the eigenvalue $T_z=0$.
For an arbitrary $T_z\neq 0$, the anticommutation relation $\{ \hat{P}\hat{T},\hat{T}_z \}=0
$
implies that $H_{0(N,\sigma,T_z)}$ and $H_{0(N,\sigma,-T_z)}$ in Eq.~\eqref{eq: non-int add symm app} are related by a unitary matrix $U_{PT}$ as
%%%%%%
\begin{align}
U_{PT} H^*_{0(N,\sigma,T_z)} U^\dagger_{PT}
= H_{0(N,\sigma,-T_z)} .
\end{align}
%%%%%%
Therefore, the topological index $\nu_{(N,\sigma)}$ characterizing the point-gap topology protected by $PT$ symmetry is given by
%%%%%%
\begin{align}
\nu_{(N,\sigma)}
&= \prod_{T_z}
\mathrm{sgn}\Big(\!
\det\big[ H_{0(N,\sigma,T_z)} - E_\mathrm{ref} \1 \big]
\Big) \nonumber \\
&=
\prod_{T_z>0}
\mathrm{sgn}\Big(
\big|
\det\big[ H_{0(N,\sigma,T_z)} - E_\mathrm{ref} \1 \big]
\big|^2
\Big)
=1
\end{align}
%%%%%%
and hence it cannot take the value $-1$.
This indicates that the system is topologically trivial in the absence of interactions.
On the other hand, once interactions are introduced, the block diagonalization of $\hat{H}=\hat{H}_0+\hat{H}_\mathrm{int}$ with respect to $\hat{T}_z$ is no longer applicable.
As a result, $\nu_{(N,\sigma)}$ can take both values $\pm1$.
From the above discussion, we conclude that the classification of the zero-dimensional point-gap topology of $H_{(N,\sigma)}$ changes due to interactions as
%%%%%%
\begin{align*}
    0 \rightarrow\mathbb{Z}_2 \quad\quad \mathrm{for}\ N+1+\sigma=0 \pmod{4}.
\end{align*}
%%%%%%

For the case $N+1+\sigma=2  \pmod{4}$, the action of $\hat{P}\hat{T}$ is also closed within the Fock-space sector labeled by $(N,\sigma)$.
In the absence of interactions, $H_{(N,\sigma)}$ can be further block-diagonalized by $\hat{T}_z$, and this Fock-space sector contains a subspace labeled by $T_z=0$, within which the $PT$ symmetry is closed.
Accordingly, $\nu_{(N,\sigma)}$ is given by
%%%%%%
\begin{align}
\nu_{(N,\sigma)}&=\prod_{T_z} \mathrm{sgn}\Big(\mathrm{det}\big[H_{0(N,\sigma,T_z)}-E_\mathrm{ref}\1 \big] \Big) \nonumber \\
&=
\mathrm{sgn}\Big(\mathrm{det}\big[H_{0(N,\sigma,0)}-E_\mathrm{ref}\1 \big]\Big)  \times \prod_{T_z>0} \mathrm{sgn}\Big(\big|\mathrm{det}\big[H_{0(N,\sigma,T_z)}-E_\mathrm{ref}\1 \big]\big|^2 \Big)
\end{align}
%%%%%%
and is therefore allowed to take both $+1$ and $-1$.
When interactions are introduced, the block diagonalization of $\hat{H}$ with respect to $\hat{T}_z$ is no longer applicable; however, the above $\mathbb{Z}_2$ topological index remains well defined even in the presence of interactions.
This suggests that the topology of $H_{(N,\sigma)}$ is preserved under interactions.
From the above discussion, the classification of the zero-dimensional point-gap topology of $H_{(N,\sigma)}$ is given by 
%%%%%%
\begin{align*}
    \mathbb{Z}_2 \rightarrow\mathbb{Z}_2 \quad\quad \mathrm{for}\ N+1+\sigma=2 \pmod{4}.
\end{align*}
%%%%%%

For the cases $N+1+\sigma=1,3 \pmod{4}$, the $PT$ symmetry is not closed within the Fock-space sector labeled by $(N,\sigma)$ [see Eq.~\eqref{eq: IEEL H-PT algebraic relation}].
Therefore, the $PT$-protected $\mathbb{Z}_2$ index is not well defined in this sector, and the zero-dimensional point-gap topology is trivial both with and without interactions.
The resulting classification is
%%%%%%
\begin{align*}
    0 \rightarrow 0 \quad\quad \mathrm{for}\ N+1+\sigma=1,3 \pmod{4}.
\end{align*}
%%%%%%

%%%%%%
\section{Properties of $PT$-symmetric non-Hermitian Hamiltonians} \label{sec: chara of PT}
%%%%%%
%%%%%%
\subsection{Characteristic-polynomial discriminants and EP2s in $PT$-symmetric systems} \label{subsec: Disc of PT sys}
%%%%%%
The discriminant of the characteristic polynomial of a matrix $H\in\mathbb{C}^{n\times n}$ is defined by
\begin{equation}
\mathrm{Disc}\big[\!\det[H-E\1_n]\big]:=\prod_{1\leq i<j\leq n}(\lambda_i-\lambda_j)^2,
\end{equation}
with $\{\lambda_i\}^n_{i=1}$ being the eigenvalues of $H$.
For a $PT$-symmetric Hamiltonian $H$, its spectrum consists of real eigenvalues as well as pairs of complex-conjugate eigenvalues with positive and negative imaginary parts.
Accordingly, we partition the index set labeling the eigenvalues into three disjoint subsets: $I_1$ labeling the $n_r$ real eigenvalues; $I_2$ labeling the $n_c$ eigenvalues with positive imaginary parts; and $I_3$ labeling their complex-conjugate counterparts.
With this partition, one obtains
%%%%%%
\begin{align}
    \mathrm{Disc}\big[\!\det[H-E\1_n]\big]&=\prod^{3}_{\alpha=1}\Big(\prod_{i<j(i,j\in I_\alpha)}(\lambda_i-\lambda_j)^2\Big)\times\prod_{1\leq \beta<\gamma\leq 3}\Big(\prod_{l\in I_\beta}\prod_{k\in I_\gamma}(\lambda_l-\lambda_k)^2\Big).
\end{align}
%%%%%%
In the first part 
on the right-hand side of the above equation, the contributions with $\alpha=2$ and $\alpha=3$
form a complex-conjugate pair.
Hence, the first part is non-negative.
In the second part, the contributions with $(\beta,\gamma)=(1,2)$
and $(\beta,\gamma)=(1,3)$ form a complex-conjugate pair,
whereas the contribution with $(\beta,\gamma)=(2,3)$ can take negative values.
Indeed, for $(\beta,\gamma)=(2,3)$, we obtain
%%%%%%
\begin{align}
    \prod_{l\in I_2}\prod_{k\in I_3}(\lambda_l-\lambda_k)^2
    &=\prod_{j\in I_2}(\lambda_j-\lambda^{*}_j)^2
    \times \prod_{l<k(l,k\in I_2)}(\lambda_l-\lambda^{*}_k)^2(\lambda_k-\lambda^{*}_l)^2 \nonumber\\
    &=(-1)^{n_c} \times\prod_{j\in I_2}(2\mathrm{Im}\lambda_j)^2\times
    \prod_{l<k(l,k\in I_2)}|\lambda_l-\lambda^{*}_k|^4 .
\end{align}
%%%%%%
Here, in going from the first line to the second line on the right-hand side of the above equation, we have used the fact that each eigenvalue $\lambda_k$ with $k\in I_3$ is in one-to-one correspondence with the complex conjugate of an eigenvalue $\lambda_k$ with $k\in I_2$ satisfying $\mathrm{Im}\lambda_k>0$.
Therefore, the sign of the discriminant of the characteristic polynomial of a $PT$-symmetric $H$ is determined by the parity of the total number $n_c$ of complex-conjugate eigenvalue pairs.

On the other hand, for $E_\mathrm{ref}\in\mathbb{R}$,
\begin{align}
    \det[H-E_\mathrm{ref}\1_n]
    &=\prod_{i\in I_1}(\lambda_i-E_\mathrm{ref})\times
    \Big|\prod_{j\in I_2}(\lambda_j-E_\mathrm{ref})\Big|^2 .
\end{align}
Thus, $\nu=\mathrm{sgn}\big(\!\det[H(\boldsymbol{\lambda})-E_\mathrm{ref}\1_n]\big)$ counts the parity of the number of real eigenvalues $\lambda_i$ satisfying $\lambda_i<E_\mathrm{ref}$.
If $E_\mathrm{ref}$ is chosen as the EP2 eigenvalue $E_{\mathrm{EP2}}$ at a parameter point $\boldsymbol{\lambda}=\boldsymbol{\lambda}_{\mathrm{EP2}}$, $\nu$ changes discontinuously upon crossing $\boldsymbol{\lambda}_{\mathrm{EP2}}$.
Correspondingly, the number of complex-conjugate eigenvalue pairs changes by one, and hence $s=\mathrm{sgn}\!\left(\mathrm{Disc}\big[
\mathrm{det}[H(\boldsymbol{\lambda})-E\1]   \big]\right)$ also exhibits a discontinuous change.

%%%%%%
\subsection{Behavior of the loss rate} \label{subsec: behavior of lossrate}
%%%%%%
For a non-Hermitian matrix $H$, we define 
$B := (H - H^\dagger)/2\ii \quad (= B^\dagger).$
Then one finds
%%%%%%
\begin{align}
    \label{eq: expected value H}
    \mathrm{Im} \langle \tilde{\psi}(t) | H | \tilde{\psi}(t) \rangle
    &= \langle \tilde{\psi}(t) | B | \tilde{\psi}(t) \rangle ,
\end{align}
%%%%%%
where
$|\tilde{\psi}(t)\rangle
= \frac{e^{-\ii t H}|\psi(0)\rangle}{\| e^{-\ii t H}|\psi(0)\rangle \|}.$
From this, we obtain the bound
%%%%%%
\begin{align}
    \label{eq: maximum expected value H}
    \left| \mathrm{Im} \langle \tilde{\psi}(t) | H | \tilde{\psi}(t) \rangle \right|
    &= \left| \langle \tilde{\psi}(t) | B | \tilde{\psi}(t) \rangle \right|
    \le \| B \| ,
\end{align}
%%%%%%
where the operator norm is defined as
$\| B \| := \max_{\| |x\rangle \| = 1} \| B |x\rangle \|.$
Therefore, for $H_{(2M,\sigma)}$, the instantaneous loss rate satisfies
%%%%%%
\begin{align}
    \label{eq: maximum Loss rate}
    |L_{(2M,\sigma)}(t)|
    &= \left|\frac{1}{\langle N_{(2M,\sigma)}(t)\rangle}\frac{d}{dt}
    \langle N_{(2M,\sigma)}(t)\rangle\right| \nonumber\\
    &=2 \left|\mathrm{Im}\frac{\Tr\!\left[ H_{(2M,\sigma)}\mathcal{P}_{(2M,\sigma)}
        \hat{\rho}(t)\mathcal{P}_{(2M,\sigma)}\right]}
        { \Tr\!\left[\mathcal{P}_{(2M,\sigma)}\hat{\rho}(t)\mathcal{P}_{(2M,\sigma)}\right] }
    \right| \nonumber\\
    &=2 \left|\mathrm{Im} \langle\tilde{\psi}_{(2M,\sigma)}(t)| H_{(2M,\sigma)}
    |\tilde{\psi}_{(2M,\sigma)}(t)\rangle\right| \nonumber\\
    &\le 2\| B_{(2M,\sigma)} \| ,
\end{align}
%%%%%%
where $B_{(2M,\sigma)}=\left(H_{(2M,\sigma)}-H^\dagger_{(2M,\sigma)}\right)/2\ii$.
Similarly, the time-averaged loss rate is bounded as
%%%%%%
\begin{align}
    |\bar{L}_{(2M,\sigma)}[T]|
    &= \left|
    \frac{1}{T}\int_0^T dt \, L_{(2M,\sigma)}(t)
    \right|
    \le 2 \| B_{(2M,\sigma)} \| .
\end{align}
%%%%%%
Here,
$|\tilde{\psi}_{(2M,\sigma)}(t)\rangle
= \frac{e^{-\ii t H_{(2M,\sigma)}}\mathcal{P}_{(2M,\sigma)}|\psi\rangle}
{\|e^{-\ii t H_{(2M,\sigma)}}\mathcal{P}_{(2M,\sigma)}|\psi\rangle\|}.$
For the bosonic toy model in Eq.~\eqref{eq: app 2-boson hamil}
and the fermionic toy model in Eq.~\eqref{eq: 2-fermi hamil},
we find $\| B_{(2,+)} \| = |\delta|$.

We assume that the non-Hermitian Hamiltonian $H$ is diagonalizable in terms of its right eigenvectors $|R_n\rangle$ and left eigenvectors $|L_n\rangle$, such that $H=\sum_nE_n|R_n\rangle\langle L_n|$.
Then, the time-evolved state is given by $|\psi(t)\rangle=$$\sum_nl_n \cdot e^{-\ii tE_n}|R_n\rangle$, where $l_n:= \langle L_n|\psi(0)\rangle$.
Using this expression, we obtain
%%%%%%
\begin{align}
\label{eq: loss rate formula 1}
\mathrm{Im}\langle\psi(t)|H|\psi(t)\rangle
&=\frac{1}{2}\frac{d}{dt}
\sum_{mn} l_m^* l_n \langle R_m|R_n\rangle
e^{-\ii t(E_n-E_m^*)} \nonumber\\
&=\frac{1}{2}\frac{d}{dt}
\sum_{mn} e^{t\beta_{mn}} Z_{mn} e^{\ii t\alpha_{mn}} .
\end{align}
%%%%%%
Here, we have defined
%%%%%%
\begin{align}
\label{eq: loss rate formula 2}
\alpha_{mn}&:= \mathrm{Re} E_m-\mathrm{Re} E_n = -\alpha_{nm},\\
\label{eq: loss rate formula 3}
\beta_{mn}&:= \mathrm{Im} E_m+\mathrm{Im} E_n = \beta_{nm},\\
\label{eq: loss rate formula 4}
Z_{mn}&:= l_m^* l_n \langle R_m|R_n\rangle = Z_{nm}^* .
\end{align}
%%%%%%
From Eqs.~\eqref{eq: loss rate formula 2}-\eqref{eq: loss rate formula 4}, the summation appearing in the second line on the right-hand side of Eq.~\eqref{eq: loss rate formula 1} can be rewritten as
%%%%%%
\begin{align}
    \label{eq: }
     \sum_{mn}e^{t\beta_{mn}}Z_{mn}e^{\ii t\alpha_{mn}}
     &=\sum_m e^{t\beta_{mm}}Z_{mm}
    +\sum_{m<n}e^{t\beta_{mn}}(Z_{mn}e^{\ii t\alpha_{mn}}+Z_{nm}e^{\ii t\alpha_{nm}})\nonumber\\
    &=\sum_m e^{t\beta_{mm}}Z_{mm}+2\sum_{m<n}e^{t\beta_{mn}}|Z_{mn}|\mathrm{cos}(\theta_{mn}+t\alpha_{mn}).
\end{align}
%%%%%%
Here we have written $Z_{mn}=|Z_{mn}|e^{\ii\theta_{mn}}$.
For a $PT$-symmetric Hamiltonian $H$, we have
$\gamma:= \max\{\mathrm{Im} E_m\}\ge 0$, and hence
$\beta_{\mathrm{max}}:= \max\{\beta_{mn}\}=2\gamma\ge 0$.
Accordingly, we define
%%%%%%
\begin{align}
G(t):=
&\sum_m e^{-t(\beta_{\mathrm{max}}-\beta_{mm})} Z_{mm} + 2\sum_{m<n} e^{-t(\beta_{\mathrm{max}}-\beta_{mn})}
|Z_{mn}|\cos\!\left(\theta_{mn}+t\alpha_{mn}\right).
\end{align}
%%%%%%
With this definition, Eq.~\eqref{eq: loss rate formula 1} can be rewritten as
%%%%%%
\begin{align}
\mathrm{Im}\langle\psi(t)|H|\psi(t)\rangle
&=\frac{1}{2}\frac{d}{dt}\!\left[e^{t\beta_{\mathrm{max}}} G(t)\right] 
=e^{t\beta_{\mathrm{max}}}\left[\gamma\,  G(t)
+\frac{1}{2}  G'(t)\right].
\end{align}
%%%%%%
For an initial state $|\psi(0)\rangle\neq\boldsymbol{0}$, we have
$\langle\psi(t)|\psi(t)\rangle = e^{t\beta_{\mathrm{max}}} G(t) > 0$.
Therefore, the loss rate $L(t)$ is given by
%%%%%%
\begin{align}
L(t)
&=-2\mathrm{Im}\langle\tilde{\psi}(t)|H|\tilde{\psi}(t)\rangle \nonumber\\
&=-2\frac{\mathrm{Im}\langle\psi(t)|H|\psi(t)\rangle}
{\langle\psi(t)|\psi(t)\rangle}
=-2\gamma-\frac{G'(t)}{G(t)} .
\end{align}
%%%%%%
For $\gamma>0$ and sufficiently large times
$t\gg(\beta_{\mathrm{max}}-\beta_{mn})^{-1}$,
the function $G(t)$ behaves asymptotically as
%%%%%%
\begin{align}
\label{eq: G_large t}
G(t)\sim
&\sum_{\substack{m\\ \beta_{mm}=\beta_{\mathrm{max}}}} Z_{mm} +2\sum_{\substack{m<n\\ \beta_{mn}=\beta_{\mathrm{max}}}}
|Z_{mn}|\cos\!\left(\theta_{mn}+t\alpha_{mn}\right).
\end{align}
%%%%%%
As a result, $L(t)$ oscillates around $-2\gamma$
(if there exists at least one pair $m\ (<n)$ that satisfies
$\beta_{mn}=\beta_{\mathrm{max}}$ and $\alpha_{mn}\neq 0$),
or converges to $-2\gamma$.

By contrast, when all eigenvalues are real, the restrictions on the summations in Eq.~\eqref{eq: G_large t} are lifted, and $L(t)$ oscillates around zero.

In both cases, since
$\frac{1}{t}\ln[G(t)/G(0)] \to 0$ as $t\to\infty$,
the long-time average satisfies $\frac{1}{T}\int^T_0dtL(t)$$\sim-2\gamma$.

%%%%%%
\section{Newton's identity} \label{sec: Newtons I}
%%%%%%
For an arbitrary square matrix ${H}\in\mathbb{C}^{n\times n}$, we define
%%%%%%
\begin{align}
\label{eq: Inverse chara poly}
P_{\mathrm{inv}}(t)
&:= \det\left[\1_n + t{H}\right] \nonumber\\
&= \prod_{i=1}^{n} (1+{\lambda}_i t)
= \sum_{k=0}^{n} e_k t^k ,
\end{align}
%%%%%%
where $\lambda_i$ ($i=1,\dots,n$) are the eigenvalues of $H$ and $e_k$ ($k=1,\dots,n$) denote the elementary symmetric polynomials $e_k=\sum_{1\leq i_1<\cdots<i_k\leq n}\lambda_{i_1}\cdots\lambda_{i_k}$ with $e_0=1$.

Let us consider the formal power-series expansion of the logarithm $\mathrm{ln}\ P_{\mathrm{inv}}(t)$:
%%%%%%
\begin{align}
    \mathrm{ln} \ P_\mathrm{inv}(t)&=\sum^n_{i=1} \ \mathrm{ln} \ (1+{\lambda}_it)\nonumber\\
    &=\sum^n_{i=1}\sum^\infty_{k=1}\frac{(-1)^{k-1}}{k}({\lambda}_it)^k\nonumber\\
    &=\sum^\infty_{k=1}\frac{(-1)^{k-1}}{k}\Tr[{H}^k]t^k.
\end{align}
%%%%%%
Taking the derivative with respect to $t$, we obtain
%%%%%%
\begin{align}
\label{eq: differential relation1}
    \frac{d}{dt}\mathrm{ln} \  P_\mathrm{inv}(t)&=\sum^\infty_{k=0}(-1)^k \ \Tr[{H}^{k+1}]t^k.
\end{align}
%%%%%%
On the other hand, differentiating Eq.~\eqref{eq: Inverse chara poly} directly yields
%%%%%%
\begin{align}
    \frac{d}{dt}P_\mathrm{inv}(t)&=\sum^{n-1}_{k=0}(k+1)e_{k+1}t^k.
\end{align}
%%%%%%
Using the identity $df/dt=f\cdot d \ (\mathrm{ln} \ f)/dt$, we also have
%%%%%%
\begin{align}
    \label{eq: differential relation2}
  \frac{d}{dt}P_\mathrm{inv}(t)&= P_\mathrm{inv}(t)\frac{d}{dt}\mathrm{ln} \  P_\mathrm{inv}(t)\nonumber\\
    &=\sum^n_{m=0}e_mt^m\sum^\infty_{l=0}(-1)^l \ \Tr[{H}^{l+1}]t^l\nonumber\\
    &=\sum^n_{m=0}\sum^\infty_{l=0}(-1)^le_m \ \Tr[{H}^{l+1}]t^{l+m}.
\end{align}
%%%%%%
By comparing the coefficients of $t^k$ in Eqs.~\eqref{eq: differential relation1} and \eqref{eq: differential relation2},
we obtain
%%%%%%
\begin{align}
    \label{eq: coeff relation}
    (k+1)&e_{k+1}={\sum_{0\leq m\leq n,l\geq0,l+m=k}}(-1)^le_m \ \Tr[{H}^{l+1}]\nonumber\\
    & \ \ \ \ \ \ \ =\sum^k_{l=0}(-1)^le_{k-l} \ \Tr[{H}^{l+1}] .
\end{align}
%%%%%%

%%%%%%
\section{Characterization of $n$- and $(n-1)$-fold EPs in $n$-band systems} \label{sec: chara multi-fold EP}
%%%%%%
As mentioned in the main text, a topological invariant allows systematic characterization of EP$3$s with $PT$ symmetry for a $4\times4$ matrix. In this section, we explain this fact for $(n-1)$-fold EPs described by an $n\times n$ matrix with/without $PT$ symmetry.
We consider $n$- and $(n-1)$-fold EPs in an arbitrary $n$-band system described by
$H\in\mathbb{C}^{n\times n}$.
To analyze degeneracies of $H$, it is convenient to work with the corresponding traceless
matrix
$\bar{H}=H-\frac{1}{n}\Tr[H]\1_n$.
The matrix $\bar{H}$ preserves the symmetries of $H$ as well as the relative
configuration of its eigenvalues.
Let $\lambda_i$ ($i=1,\dots,n$) be the eigenvalues of $H$.
Then
%%%%%%
\begin{align}
\label{eq: general chara poly}
   P_{\bar{H}}(E)&:= \det[\bar{H}-E\1_n] \nonumber\\
   &=\prod^n_{i=1}(\bar{\lambda}_i-E)
   =\sum^n_{k=0}(-1)^k\bar{e}_{n-k}E^k
\end{align}
%%%%%%
holds, where
$\bar{\lambda}_i:=\lambda_i-\frac{1}{n}\Tr[H]$.
Here, $\bar{e}_k$ denotes the $k$-th elementary symmetric polynomial of the eigenvalues
$\bar{\lambda}_i$ of $\bar{H}$, defined as
$\bar{e}_k:= \sum_{1\leq i_1<\cdots<i_k\leq n}
\bar{\lambda}_{i_1}\cdots\bar{\lambda}_{i_k}$,
with $\bar{e}_0:= 1$.
These coefficients can be expressed in terms of the traces $\Tr[\bar{H}^k]$ via Newton’s identities,
%%%%%%
\begin{align}
\label{eq: Newtons identity}
(k+1)\bar{e}_{k+1}
=\sum_{l=0}^k (-1)^l \bar{e}_{k-l}\Tr[\bar{H}^{\,l+1}],
\end{align}
%%%%%%
whose derivation is provided in Sec.~\ref{sec: Newtons I}.

Prior to the characterization of $(n-1)$-fold EPs, we discuss $n$-fold EPs.
If $H$ has an $n$-fold degenerate eigenvalue $r\in\mathbb{C}$, all eigenvalues of $\bar{H}$ must vanish. In this case, the characteristic polynomial of the corresponding traceless matrix $\bar{H}$ is given by
%%%%%%
\begin{align}
\label{eq: n-fold chara poly}
P_{\bar{H}}(E)&= (-E)^n .
\end{align}
%%%%%%
Since $\bar{e}_1=\Tr[\bar{H}]=0$, Eqs.~\eqref{eq: general chara poly} and \eqref{eq: n-fold chara poly} imply that, if $H$ has an $n$-fold degenerate eigenvalue $r\in\mathbb{C}$, the coefficients $\bar{e}_k$ vanish for $k=2,\dots,n$. Therefore, imposing these $n-1$ independent conditions indicates that $H$ exhibits an EP$n$ with eigenvalue $r=\frac{1}{n}\Tr[H]$.
This observation also clarifies the codimension of an EP$n$. In a generic $n$-band system, the conditions $\bar{e}_k=0$ $(k=2,\dots,n)$ are complex constraints, yielding a codimension $2(n-1)$. In contrast, in the presence of $PT$ symmetry, these coefficients are restricted to be real, so that each condition reduces to a single real constraint. As a result, the codimension is reduced to $n-1$ under $PT$ symmetry.

Next, we consider the characterization of $(n-1)$-fold EPs. 
When $H$ has an $(n-1)$-fold degenerate eigenvalue $r$ and a single eigenvalue
$s\ (\neq r)$,
for the corresponding traceless matrix
$\bar{H}$,
%%%%%%
\begin{align}
    \label{eq: n-1-fold trace constraint}
&\Tr[H]=(n-1)r+s\Rightarrow\Tr[\bar{H}]=(n-1)\bar{r}+\bar{s}=0
\end{align}
%%%%%%
and
%%%%%%
\begin{align}
\label{eq: n-1-fold det constraint}
&\det[H]=r^{n-1}s\Rightarrow\det[\bar{H}]=(1-n)\bar{r}^n
\end{align}
%%%%%%
hold. Here we define $\bar{r}:= r-\frac{1}{n}\Tr[H]$ and $\bar{s}:= s-\frac{1}{n}\Tr[H]$.
Taking these relations into account, we obtain
%%%%%%
\begin{align}
\label{eq: n-1-fold chara poly}
   P_{\bar{H}}(E)&=(\bar{r}-E)^{n-1}(\bar{s}-E) \nonumber\\
   &=\sum^{n}_{k=0}(-1)^{n-k} \binom{n}{k} \left(1-k\right)\bar{r}^{k}E^{n-k}.
\end{align}
%%%%%%
Thus, when $H$ has an $(n-1)$-fold eigenvalue $r$ and a single eigenvalue
$s\ (\neq r)$, each coefficient $\bar{e}_k$ $(k=2,\dots,n-1)$ satisfies
$\bar{e}_k=\binom{n}{k}(1-k)\bar{r}^{k}$ [see Eqs.~\eqref{eq: general chara poly} and~\eqref{eq: n-1-fold chara poly}].
Here, $\bar{r}$ is given by
$\bar{r}=\left(\det[\bar{H}]/(1-n)\right)^{1/n}$
[see Eq.~\eqref{eq: n-1-fold det constraint}].
Combining these relations, we obtain
%%%%%%
\begin{align}
\label{eq: EPn-1 condition 1}
  &\left(\frac{\bar{e}_k}{ \binom{n}{k}(1-k)}\right)^n=\left(\frac{\det[\bar{H}]}{1-n}\right)^k,
\end{align}
%%%%%%
for $k=2,\dots,n-1$.
Since $s\neq r$, it is necessary that $\det[\bar{H}]\neq0$.
Therefore, the existence of an $(n-1)$-fold EP is confirmed when
$\det[\bar{H}]\neq0$ and the above $n-2$ conditions in
Eq.~\eqref{eq: EPn-1 condition 1} are satisfied.
Accordingly, the codimension of an $(n-1)$-fold EP for a generic $n$-band system
is $2(n-2)$, while it is reduced to $n-2$ in the presence of $PT$ symmetry.
Moreover, the $(n-1)$-fold degenerate eigenvalue $r$ of $H$ is given by
%%%%%%
\begin{align}
    \label{eq: EPn-1 eigen}
    r&=\bar{r}+\frac{1}{n}\Tr[H]
    =\left(\frac{\det\left[H-\frac{1}{n}\Tr[H]\1_n\right]}{1-n}\right)^{\frac{1}{n}}+\frac{1}{n}\Tr[H].
\end{align}
%%%%%%

Using the coefficient constraints of the characteristic polynomial in
Eq.~\eqref{eq: EPn-1 condition 1}, we can introduce the
coefficient winding number $W_c$, in analogy with the resultant
winding number $W_r$.

For a generic $n$-band system with $n\ \geq3$, 
an $(n-1)$-fold EP is characterized by the vanishing of the following $2(n-2)$-dimensional vector:
%%%%%%
\begin{align}
\boldsymbol{C}(\boldsymbol{\lambda})
=\Big(
\mathrm{Re}\, c_1,
\mathrm{Im}\, c_1,
\dots,
\mathrm{Re}\, c_{n-2},
\mathrm{Im}\, c_{n-2}
\Big)^{T}.
\end{align}
%%%%%%
Here, each component $c_k(\boldsymbol{\lambda})$ $(k=1,\dots,n-2)$ is defined as
%%%%%%
\begin{align}
\label{eq: component define}
  c_k(\boldsymbol{\lambda})&=\left(\frac{\bar{e}_{k+1}(\boldsymbol{\lambda})}{ \binom{n}{k+1}(-k)}\right)^n-\left(\frac{\det[\bar{H}(\boldsymbol{\lambda})]}{1-n}\right)^{k+1}.
\end{align}
%%%%%%
Let $\boldsymbol{\lambda}_0 \in X$ denote a parameter point corresponding to an $(n-1)$-fold EP in $2(n-2)$-dimensional parameter space $X$.
For parameter points $\boldsymbol{p} \in S^{2(n-2)-1} \subset X \setminus \{\boldsymbol{\lambda}_0\}$ surrounding $\boldsymbol{\lambda}_0$, one can introduce a mapping
between $(2n-5)$-spheres,
$\boldsymbol{n}=\boldsymbol{C}/\|\boldsymbol{C}\|:S^{2n-5}\rightarrow S^{2n-5}$.
The homotopy class of this mapping is characterized by
$\pi_{2n-5}(S^{2n-5}) = \mathbb{Z}$
and thus takes integer values, which are given by the winding number
%%%%%%
\begin{align}
    W_c
    = \frac{\epsilon^{i_1\cdots i_{2n-5}}}{2\pi^{n-2}/(n-3)!}
    \int d^{2n-5}\boldsymbol{p} \,
    f_{i_1\cdots i_{2n-5}} .
\end{align}
%%%%%%
Here, $\epsilon^{i_1\cdots i_{2n-5}}$ is the totally antisymmetric tensor satisfying
$\epsilon^{12\cdots (2n-5)} = 1$,
and
$f_{i_1\cdots i_{2n-5}}
= n_{i_1}\partial_1 n_{i_2}\partial_2 n_{i_3}\cdots \partial_{2n-5} n_{2n-4}$.

Moreover, for an $(n-1)$-fold EP at $\boldsymbol{\lambda}=\boldsymbol{\lambda}_0$ in $(n-2)$-dimensional parameter space $X'$ protected by $PT$ symmetry, an $(n-2)$-component vector
$\boldsymbol{C}(\boldsymbol{\lambda})
=\big(c_1(\boldsymbol{\lambda}),\dots,c_{n-2}(\boldsymbol{\lambda})\big)^{T}$
vanishes.
Therefore, for parameter points
$\boldsymbol{p}\in S^{n-3}\subset X'\setminus\{\boldsymbol{\lambda}_0\}$
surrounding $\boldsymbol{\lambda}_0$, one can introduce a mapping between $(n-3)$-spheres,
$\boldsymbol{n}=\boldsymbol{C}/\|\boldsymbol{C}\|:S^{n-3}\rightarrow S^{n-3}$,
from which the winding number
%%%%%%
\begin{align}
\label{eq: PT coeff wind}
    W_{c}
    =\frac{\epsilon^{i_1\cdots i_{n-3}}}{A_{n-3}}
    \int d^{n-3}\boldsymbol{p}\,
    f_{i_1\cdots i_{n-3}}
\end{align}
%%%%%%
is obtained.
Here, for a nonnegative integer $m$,
%%%%%%
\begin{align}
    A_{n-3}=
    \begin{cases}
        2\pi^{m+1}/m!
        \qquad\qquad\ \ \ \ \ (n=2(m+2)),
        \\
        2^{2m+1}\pi^{m}m!/(2m)!
        \qquad (n=2m+3),
    \end{cases}
\end{align}
%%%%%%
which corresponds to the surface area of the $(n-3)$-dimensional unit sphere.

%%%%%%
\section{
Numerical data for the bosonic toy model
} \label{sec: nume data for boson}
%%%%%%

%%%%%%%%%%%%%%%%%%%%%%%%%
\begin{figure}[!t]
\begin{minipage}{1\hsize}
\begin{center}
    \includegraphics[width=\linewidth,height=2.9cm,keepaspectratio]{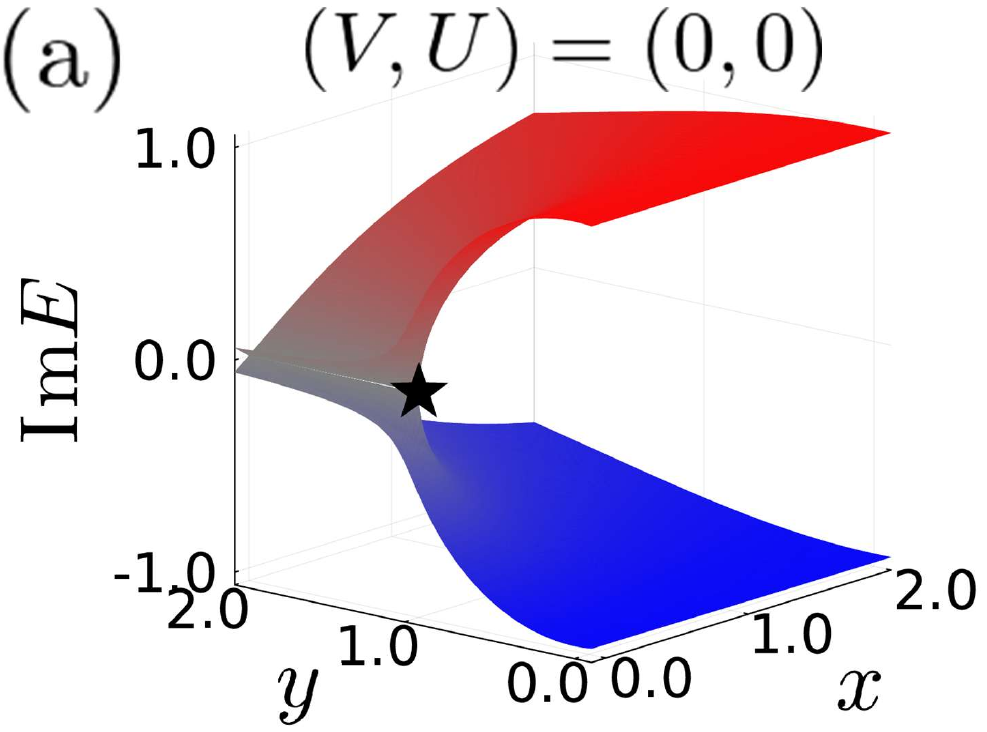}
    \includegraphics[width=\linewidth,height=3cm,keepaspectratio]{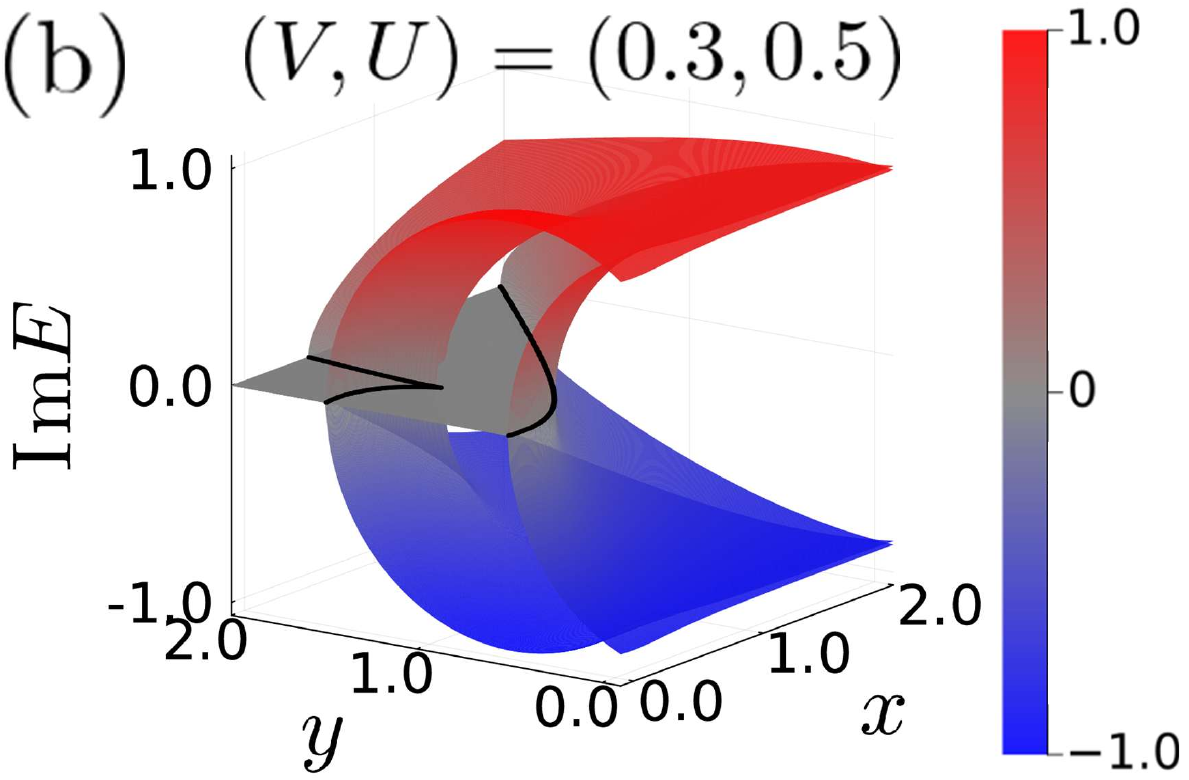}
\end{center}
\end{minipage}
\caption{
(a) [(b)]: $x$-$y$ dependence of the imaginary parts of the complex eigenvalues of $H_{(2,+)}$ for $(w,\delta)=(0,1)$ with $V=U=0$ [$(V,U)=(0.3,0.5)$]. In panel (a), the black star at $(x,y)=(0,\delta)$ represents an EP2, wheres in panel (b), black lines represent EL2s.
}
\label{fig:IEEL boson ImExy}
\end{figure}
%%%%%%%%%%%%%%%%%%%%%%%%%
In the main text, we demonstrated the emergence of interaction-enabled EL2s using the topological index $s_{(2,+)}(\boldsymbol{\lambda})$, defined as $s_{(2,+)}(\boldsymbol{\lambda})=\mathrm{sgn}\Big(\mathrm{Disc}\big[\!\det[H_{(2,+)}(\boldsymbol{\lambda})]\big]\Big)$. For clarity, we rewrite $H_{(2,+)}(\boldsymbol{\lambda})$ explicitly as
%%%%%%
\begin{align}
\label{eq: app 2-boson hamil}
H_{(2,+)}(x,y)&=\begin{pmatrix}
        H_{0(2,+,1)}(x,y) &  \\ & H_{0(2,+,-1)}(x,y)
    \end{pmatrix}+H_{\mathrm{int}(2,+)}\nonumber\\
    &=w\1_4+\begin{pmatrix}
        x-\ii\delta & y & &\\
        y & -x+\ii\delta & &\\
        & &-x-\ii\delta& y\\
        & & y & x+\ii\delta
    \end{pmatrix}+\begin{pmatrix}
  &  & V&U \\
  &  & U & V \\
V & U & & \\
 U & V& & \\
\end{pmatrix},
\end{align}
%%%%%%
where omitted entries are zero.
Figure~\ref{fig:IEEL boson ImExy} shows the $x$–$y$ dependence of the imaginary parts of the complex eigenvalues of $H_{(2,+)}(\boldsymbol{\lambda})$.
As shown in Fig.~\ref{fig:IEEL boson ImExy}(a), in the absence of interactions, no one-dimensional regions of degenerate eigenvalues appear.
In contrast, as shown in Fig.~\ref{fig:IEEL boson ImExy}(b), the presence of interactions gives rise to one-dimensional regions, indicated by the black lines, where the imaginary parts of the eigenvalues become degenerate.
In these regions, the real parts of the eigenvalues are simultaneously degenerate.
These results provide a direct visualization of the emergence of interaction-enabled EL2s.

%%%%%%%%%%%%%%%%%%%%%%%%%
\begin{figure}[!t]
\begin{minipage}{1\hsize}
\begin{center}
    \includegraphics[width=\linewidth,height=3.5cm,keepaspectratio]{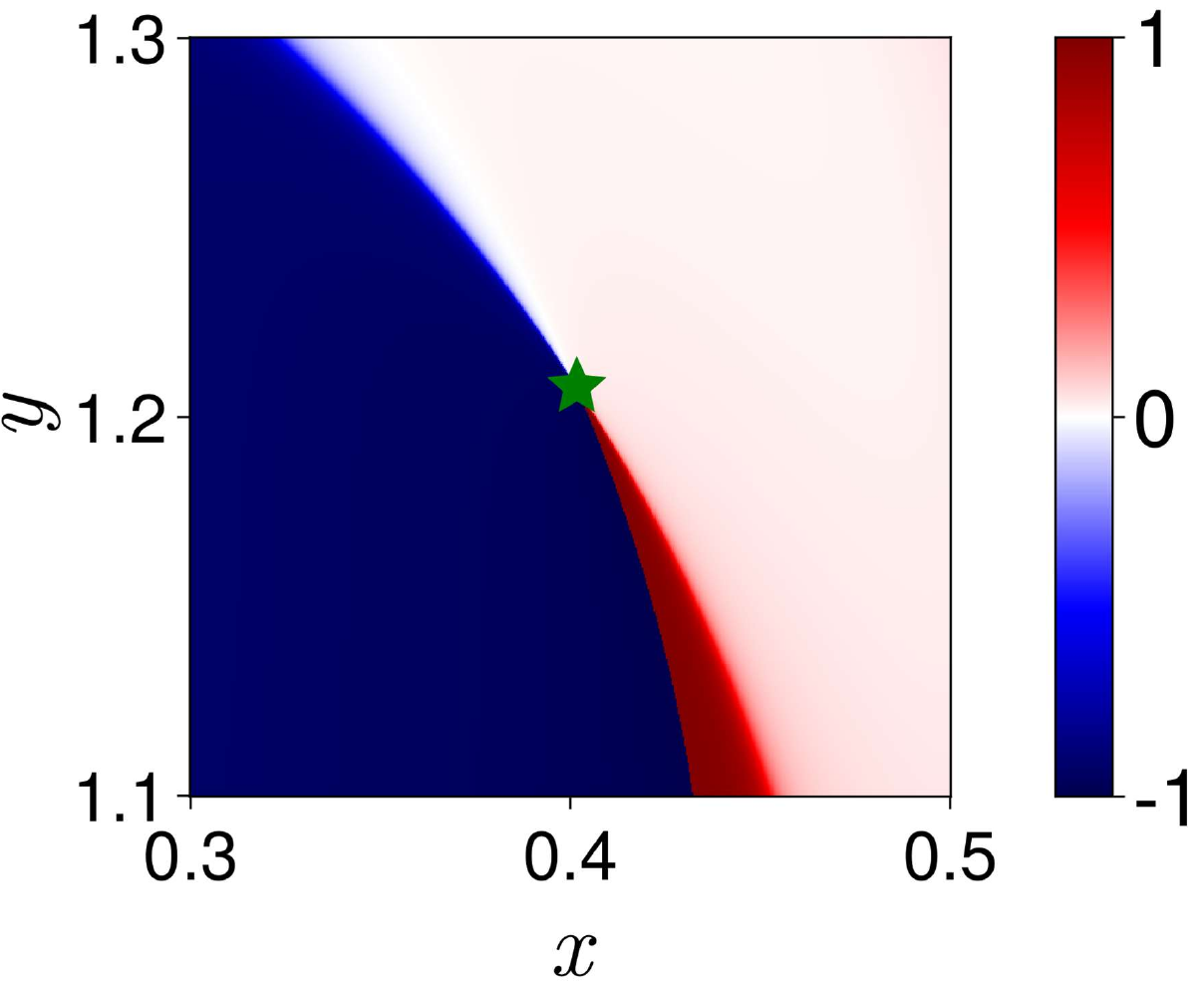}
\end{center}
\end{minipage}
\caption{
$x$–$y$ dependence of the principal value of the argument of $Z_c$ divided by $\pi$ for $(w,\delta,V,U)=(0,1,0.3,0.5)$. The green stars indicate $\boldsymbol{\lambda}_{\mathrm{EP3}}\sim(0.402,1.208)$.
}
\label{fig: IEEP3 Arg}
\end{figure}
%%%%%%%%%%%%%%%%%%%%%%%%%

In Fig.~\ref{fig:IEEL boson ImExy}(b), one can find a cusp structure of the interaction-enabled EL2 
which corresponds to the interaction-enabled EP3 discussed in the main text. 
While the resultant winding number can characterize the interaction-enabled EP3,
we characterize it using an alternative topological invariant $W_c$ introduced in Sec.~\ref{sec: chara multi-fold EP}.
Since $H_{(2,+)}(\boldsymbol{\lambda})$ is a $PT$-symmetric four-band model, a topological invariant for the interaction-enabled EP3s can be introduced as
%%%%%%
\begin{align}
\label{eq: new EP3 inv}
W_c=\oint\frac{d\boldsymbol{\lambda}}{2\pi \ii}\cdot\partial_{\boldsymbol{\lambda}} \ln Z_c(\boldsymbol{\lambda}),
\end{align}
%%%%%%
where $Z_c(\boldsymbol{\lambda})=c_1(\boldsymbol{\lambda})+\ii c_2(\boldsymbol{\lambda})$, and $c_k(\boldsymbol{\lambda})$ ($k=1,2$) are defined in Eq.~\eqref{eq: component define}.
This invariant $W_c$ is equivalent to setting $n=4$ in Eq.~\eqref{eq: PT coeff wind}.
Figure~\ref{fig: IEEP3 Arg} shows the $x$–$y$ dependence of the principal value of the argument of $Z_c$.
When the parameters encircle $\boldsymbol{\lambda}{\mathrm{EP3}}$ counterclockwise, we obtain the winding number $W_c=-1$.
Namely, the zeros of $\boldsymbol{C}(\boldsymbol{\lambda})=\big(c_1(\boldsymbol{\lambda}),c_2(\boldsymbol{\lambda})\big)^T$ enable the detection of interaction-enabled EP3s of $H_{(2,+)}(\boldsymbol{\lambda})$ over the entire two-dimensional parameter space, independent of the choice of $E_{\mathrm{EP3}}$.

%%%%%%
\section{Interaction-enabled EL2s and EP3s in fermionic systems}
\label{sec: results of fermionic sys}
%%%%%%
In the main text, we have
revealed the existence of interaction-enabled EL2s and EP3s in bosonic systems. 
Our framework can be applied to fermionic systems as well.
In this section, we show, using a toy model, that interaction-enabled EL2s and EP3s also emerge in fermionic systems with charge U(1), pseudo-spin-parity, and $PT$ symmetries.

We consider a two-component fermionic toy model:
%%%%%%
\begin{align}
\label{eq: IEEL fermionic toy model}
&\hat{H}(x,y)=\hat{H}_{0\mathrm{A}}(x,y)+\hat{H}_{0\mathrm{B}}(x,y)+\hat{H}_{\mathrm{int}},\\
    &\hat{H}_{0\tau}=\hat{\Psi}^\dagger_{f,\tau}h_{\tau}(x,y)\hat{\Psi}_{f,\tau} \  \ (\tau=\mathrm{A},\mathrm{B}),\\
    &\hat{H}_{\mathrm{int}}=V\sum^3_{i=1}(\hat{f}^{\dagger}_{i,\mathrm{A}} \hat{f}_{i,\mathrm{B}}\hat{f}^{\dagger}_{i+1,\mathrm{A}}\hat{f}_{i+1,\mathrm{B}}+\mathrm{h.c.}),
\end{align}
%%%%%%
where real parameters $x$ and $y$ describe a two-dimensional parameter space. Here, $\hat{\Psi}^\dagger_{f,\tau}=\big(\hat{f}^\dagger_{1,\tau},\ \hat{f}^\dagger_{2,\tau},\ \hat{f}^\dagger_{3,\tau}\big)$, where $\hat{f}^\dagger_{i,\tau}\ (\hat{f}_{i,\tau})$ is the creation (annihilation) operator of a fermion with component $\tau\ (=\mathrm{A,B})$ on site $i\ (=1,2,3)$. In the interaction term, $V$ is a real parameter representing the interaction strength, and we identify $i+1:=1$ when $i=3$.
The first-quantized Hamiltonian $h_\tau(x,y)$ is given by
%%%%%%
\begin{align}
     &h_{\tau}(x,y)=\begin{pmatrix}
        (-1)^{\delta_{\mathrm{B},\tau}}x-\ii\delta & y & y\\
        y &  0 & y\\
        y & y & -(-1)^{\delta_{\mathrm{B},\tau}}x+\ii\delta
    \end{pmatrix}.
\end{align}
%%%%%%
$\hat{H}(x,y)$ preserves charge U(1), pseudo-spin-parity, and $PT$ symmetries [Eqs.~\eqref{eq: app IEEL c-U1 symm}–\eqref{eq: app IEEL PT symm}].
The $PT$ operator is defined as
%%%%%%
\begin{align}
      &\hat{P}\hat{T}=\mathrm{exp}\left[\ii\frac{\pi}{2}\sum_{i,j}(p_{ij}\hat{f}^\dagger_{i,\mathrm{A}}\hat{f}_{j,\mathrm{B}}+p^T_{ij}\hat{f}^\dagger_{i,\mathrm{B}}\hat{f}_{j,\mathrm{A}})\right]\mathcal{K},
\end{align}
%%%%%%
where $p=\begin{pmatrix} 0 & 0 & 1\\0 &1 & 0\\1 & 0 & 0 \end{pmatrix}$
and $\mathcal{K}$ denotes the complex conjugation operator.

%%%%%%%%%%%%%%%%%%%%%%%%%
\begin{figure}[!b]
\begin{minipage}{1\hsize}
\begin{center}
    \includegraphics[width=\linewidth,height=3.9cm,keepaspectratio]{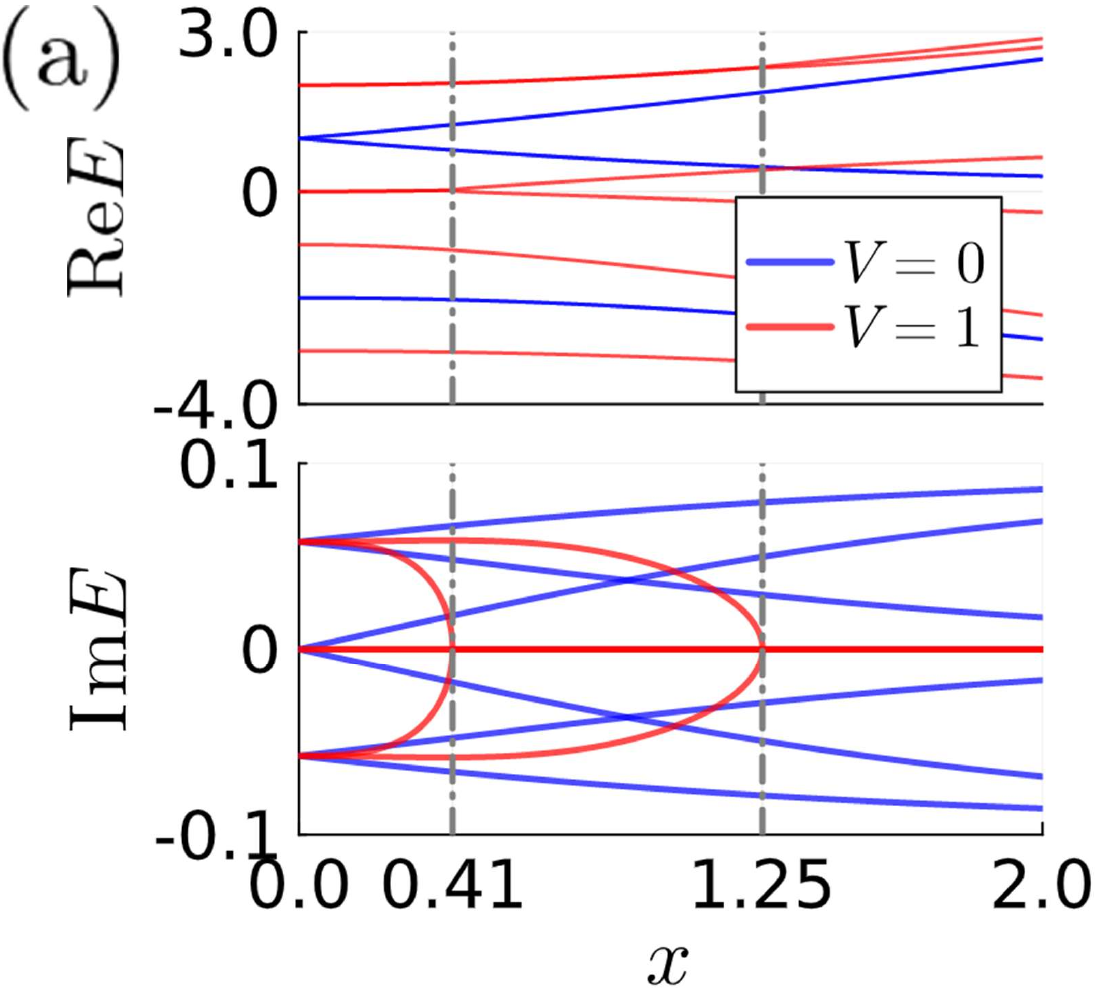}
    \includegraphics[width=\linewidth,height=3.9cm,keepaspectratio]{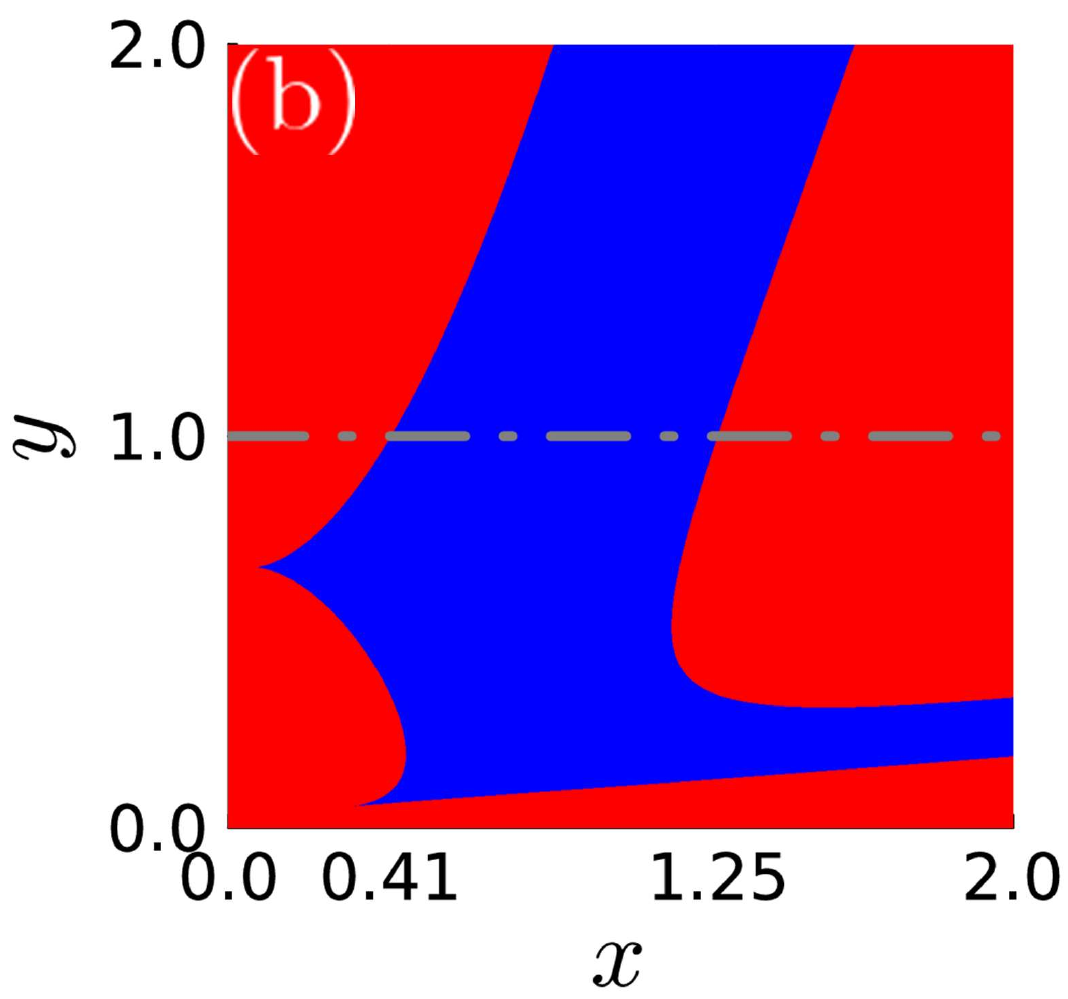}
\end{center}
\end{minipage}
\caption{
(a): $x$-dependence of the eigenvalues of 
$H_{(2,+)}(x,y=1)$ 
for $\delta=0.1$. The blue and red solid lines correspond to $V=0$ and $V=1$, respectively.
(b): The $\mathbb{Z}_2$ topological index $s_{(2,+)}(x,y)$ for $\delta=0.1$ and $V=1$. The blue and red regions correspond to $s=-1$ and $s=+1$, respectively.
}
\label{fig:IEEL fermion Ex_s}
\end{figure}
%%%%%%%%%%%%%%%%%%%%%%%%%

%%%%%%%%%%%%%%%%%%%%%%%%%
\begin{figure}[!t]
\begin{minipage}{1\hsize}
\begin{center}
    \includegraphics[width=\linewidth,height=2.85cm,keepaspectratio]{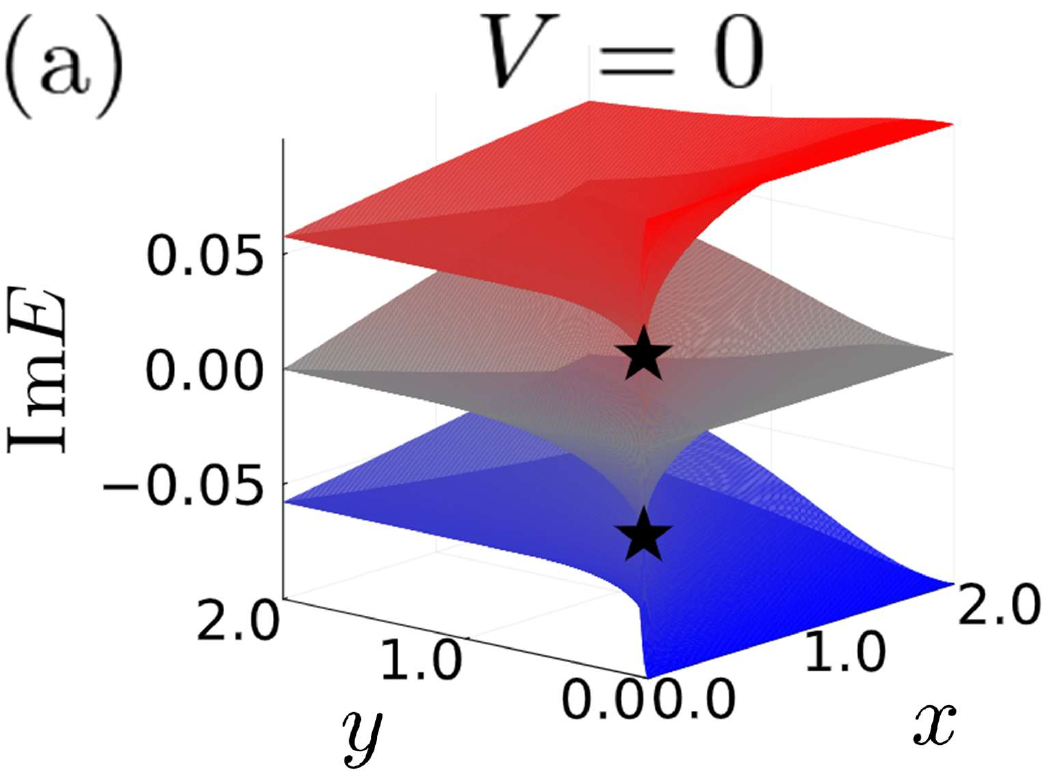}
    \includegraphics[width=\linewidth,height=2.85cm,keepaspectratio]{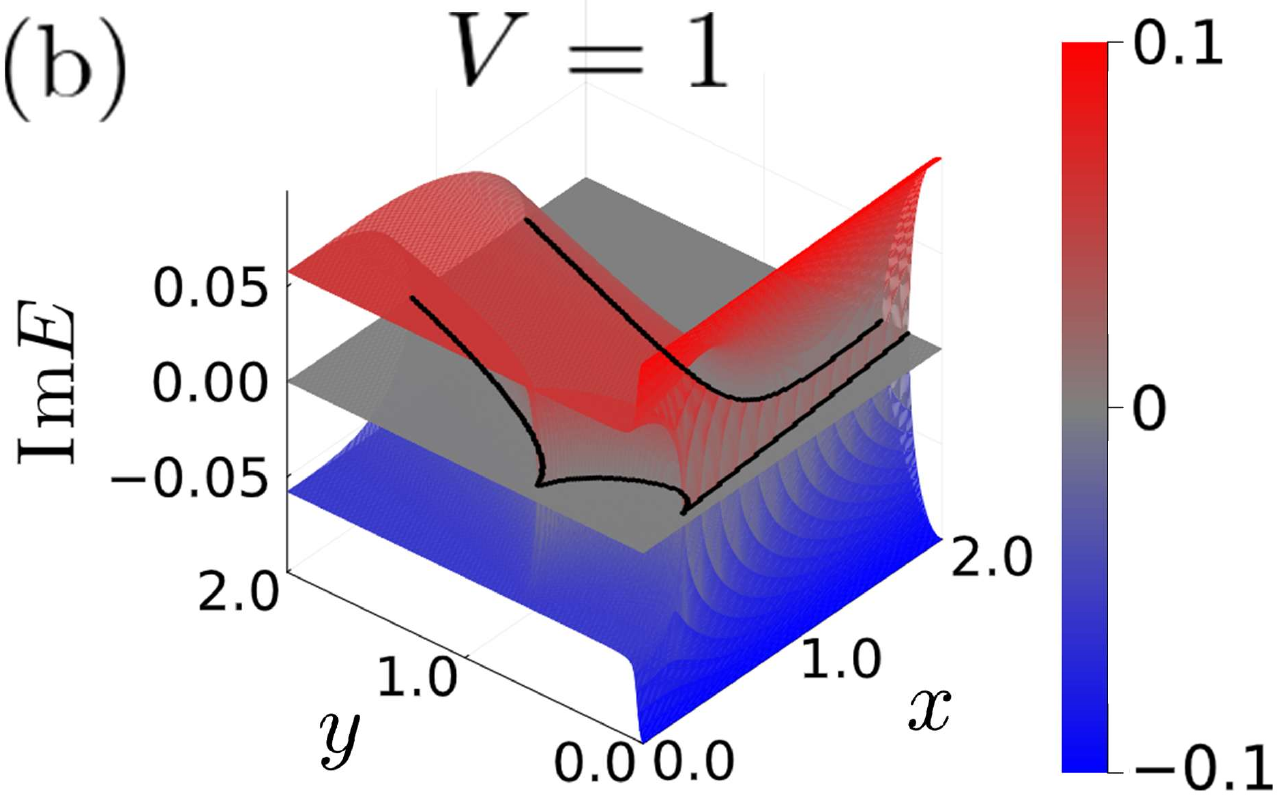}
\end{center}
\end{minipage}
\caption{
(a) [(b)]: $x$-$y$ dependence of the imaginary parts of the complex eigenvalues of $H_{(2,+)}$ for $\delta=0.1$ with $V=0$ [$V=1$]. In panel (a), the black stars at $(x.y,\mathrm{Im} E)=\delta\big(\sqrt{7-4\sqrt{3}},\frac{2}{3}\sqrt{2\sqrt{3}-3},\pm\frac{\sqrt{3}}{3}\sqrt{2\sqrt{3}-3}\big)$ each represent an EP2, wheres in panel (b), black lines represent EL2s.
}
\label{fig:IEEL fermion ImExy}
\end{figure}
%%%%%%%%%%%%%%%%%%%%%%%%%

We focus on the Fock-space sector with $\big(N,(-1)^{N_\mathrm{B}}\big)=(2,+)$, satisfying $N+1+\sigma = 0 \pmod{4}$.
The Hamiltonian $H_{(2,+)}$ which acts closedly within this subspace is given by,
%%%%%%
\begin{align}
    \label{eq: 2-fermi hamil}
    &H_{(2,+)}=\begin{pmatrix}
        H_{0(2,+,T_z=+1)} & O \\O & H_{0(2,+,T_z=-1)}
    \end{pmatrix}
     +V
     \begin{pmatrix}
        O & \1_3\\\1_3 & O
    \end{pmatrix},
    \\
   &H_{0(2,+,T_z)}=\begin{pmatrix}
        (-1)^{\delta_{T_z,-1}}x-\ii\delta & -y & -y\\
        -y & \ii\delta-(-1)^{\delta_{T_z,-1}}x & -y\\
        -y & -y & 0
    \end{pmatrix}.
\end{align}
%%%%%%
The basis is chosen as
%%%%%%
\begin{align}
\label{eq: 2-fermi basis}
\Big(|\mathrm{A}1,\mathrm{A}2\rangle,|\mathrm{A}2,\mathrm{A}3\rangle,|\mathrm{A}3,\mathrm{A}1\rangle, |\mathrm{B}1,\mathrm{B}2\rangle,|\mathrm{B}2,\mathrm{B}3\rangle,|\mathrm{B}3,\mathrm{B}1\rangle\Big),
\end{align}
%%%%%%
where the basis states are defined as $|\tau i,\tau j\rangle=\hat{f}^\dagger_{i,\tau}\hat{f}^\dagger_{j,\tau}|0\rangle$.
Since the relations $H_{(2,+)}(-x,y)=SH_{(2,+)}(x,y)S^{-1}$ with $ S=\tau_1\otimes\1_3$ and $H_{(2,+)}(x,-y)=-R\mathcal{K}H_{(2,+)}(x,y)(R\mathcal{K})^{-1}$ with $R=\tau_2\otimes\1_3$ hold, we focus our analysis on the first quadrant of the $x$-$y$ plane.

We choose $y=1$ as an illustrative example and plot the $x$-dependence of the complex eigenvalues of $H_{(2,+)}(x,y=1.0)$ in Fig.~\ref{fig:IEEL fermion Ex_s}(a).
In contrast to the case for $V=0$, both the real and imaginary parts of the eigenvalues become degenerate at $y\sim 0.41$ and $y\sim 1.25$ for $V=1$, indicating the emergence of interaction-enabled EP2s.
Correspondingly, the $\mathbb{Z}_2$ index
$s_{(2,+)}(x,y=1.0)$ jumps at these EP2s [see Fig.~\ref{fig:IEEL fermion Ex_s}(b)], characterizing their topology. 
Moreover, Fig.~\ref{fig:IEEL fermion Ex_s}(b) shows that these EP2s extend to other values of $y$, indicating the emergence of interaction-enabled EL2s in the $x$-$y$ plane (see also Fig.~\ref{fig:IEEL fermion ImExy}).

%%%%%%%%%%%%%%%%%%%%%%%%%
\begin{figure}[!t]
\begin{minipage}{1\hsize}
\begin{center}
    \includegraphics[width=\linewidth,height=3.5cm,keepaspectratio]{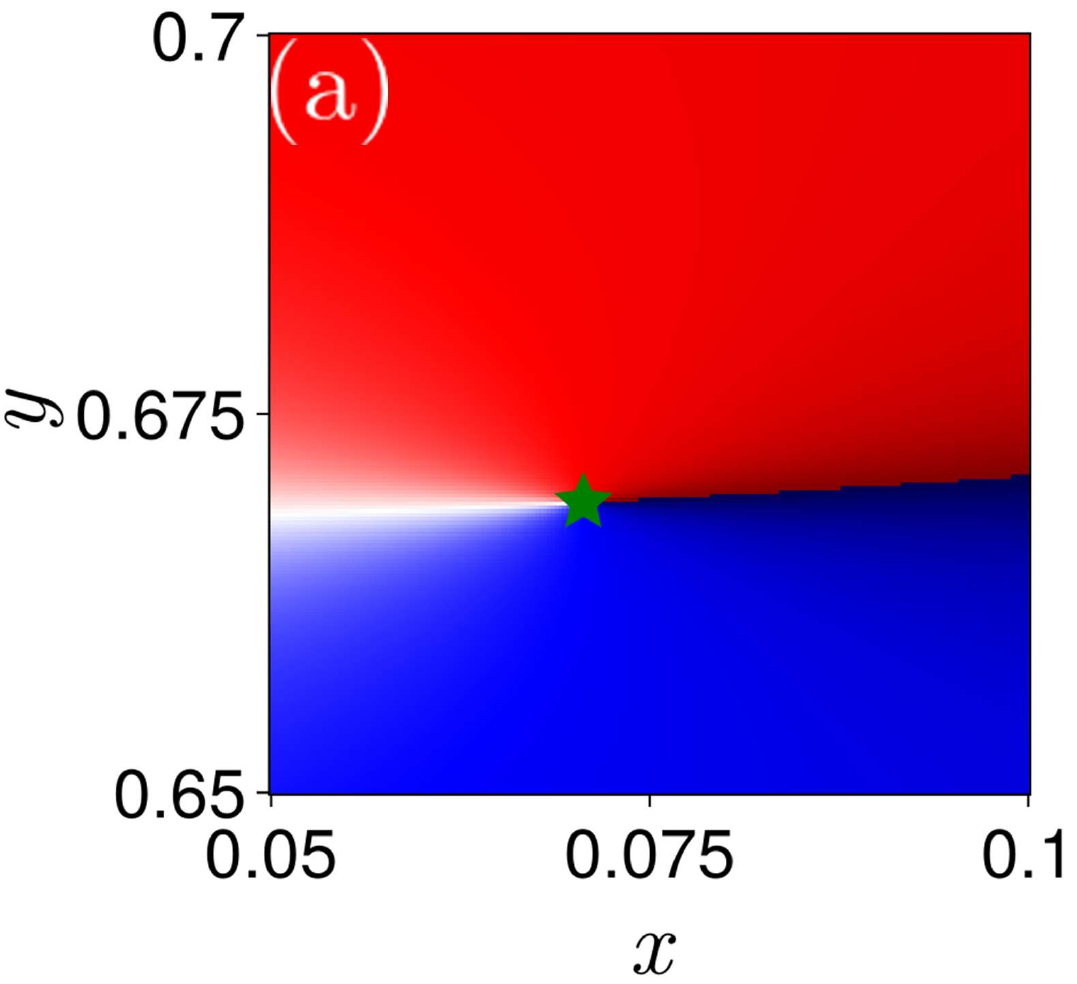}
    \includegraphics[width=\linewidth,height=3.5cm,keepaspectratio]{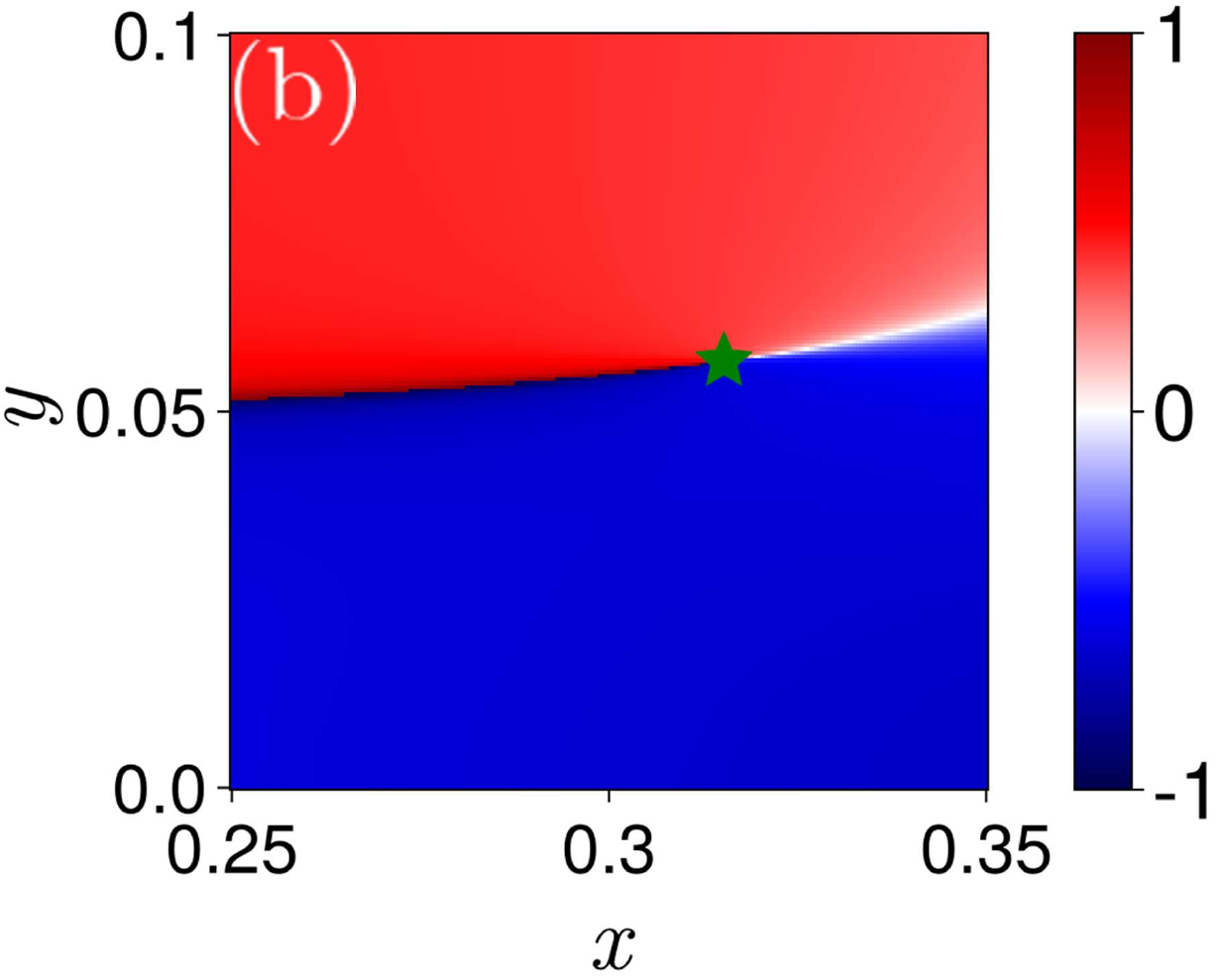}
\end{center}
\end{minipage}
\caption{
(a) [(b)]: $x$–$y$ dependence of the principal value of $Z_r$ for 
$(\delta,V)=(0.1,1)$ 
with $E_{\mathrm{ref},1}\sim-1.033$ [$E_{\mathrm{ref},2}\sim-0.331$] divided by $\pi$. 
The green star indicates $\boldsymbol{\lambda}_{1}\sim(0.071,0.669)$ [$\boldsymbol{\lambda}_{2}\sim(0.315,0.057)$].
}
\label{fig: fIEEP3 Arg}
\end{figure}
%%%%%%%%%%%%%%%%%%%%%%%%%
Furthermore, the two cusps on the interaction-enabled EL2s observed in Fig.~\ref{fig:IEEL fermion Ex_s}(b) at
$\boldsymbol{\lambda}_1\sim(0.071,0.669)$ and $\boldsymbol{\lambda}_2\sim(0.315,0.057)$
are identified as interaction-enabled EP3s with $\mathbb{Z}$ topology through the resultant winding number.
Indeed, when the parameters $(x,y)$ encircle the cusps $\boldsymbol{\lambda}_1$ and $\boldsymbol{\lambda}_2$ counterclockwise,
the resultant winding number takes the values $-1$ and $+1$, respectively (see Fig.~\ref{fig: fIEEP3 Arg}).

The existence of interaction-enabled EL2s
protected by $PT$ symmetry is also reflected in the behavior of the loss rate in fermionic systems. 
We set the initial state as $|\psi(0)\rangle=\frac{1}{\sqrt{2}}\big(|\mathrm{A}2,\mathrm{A}3\rangle+|\mathrm{B}1,\mathrm{B}2\rangle\big)$ 
and discuss the behavior of the loss rate at $y=1$, the same value as in Fig.~\ref{fig:IEEL fermion Ex_s}(a).
%
%%%%%%%%%%%%%%%%%%%%%%%%%
\begin{figure}[!t]
\begin{minipage}{1\hsize}
\begin{center}
    \includegraphics[width=0.49\linewidth,height=3.6cm,keepaspectratio]{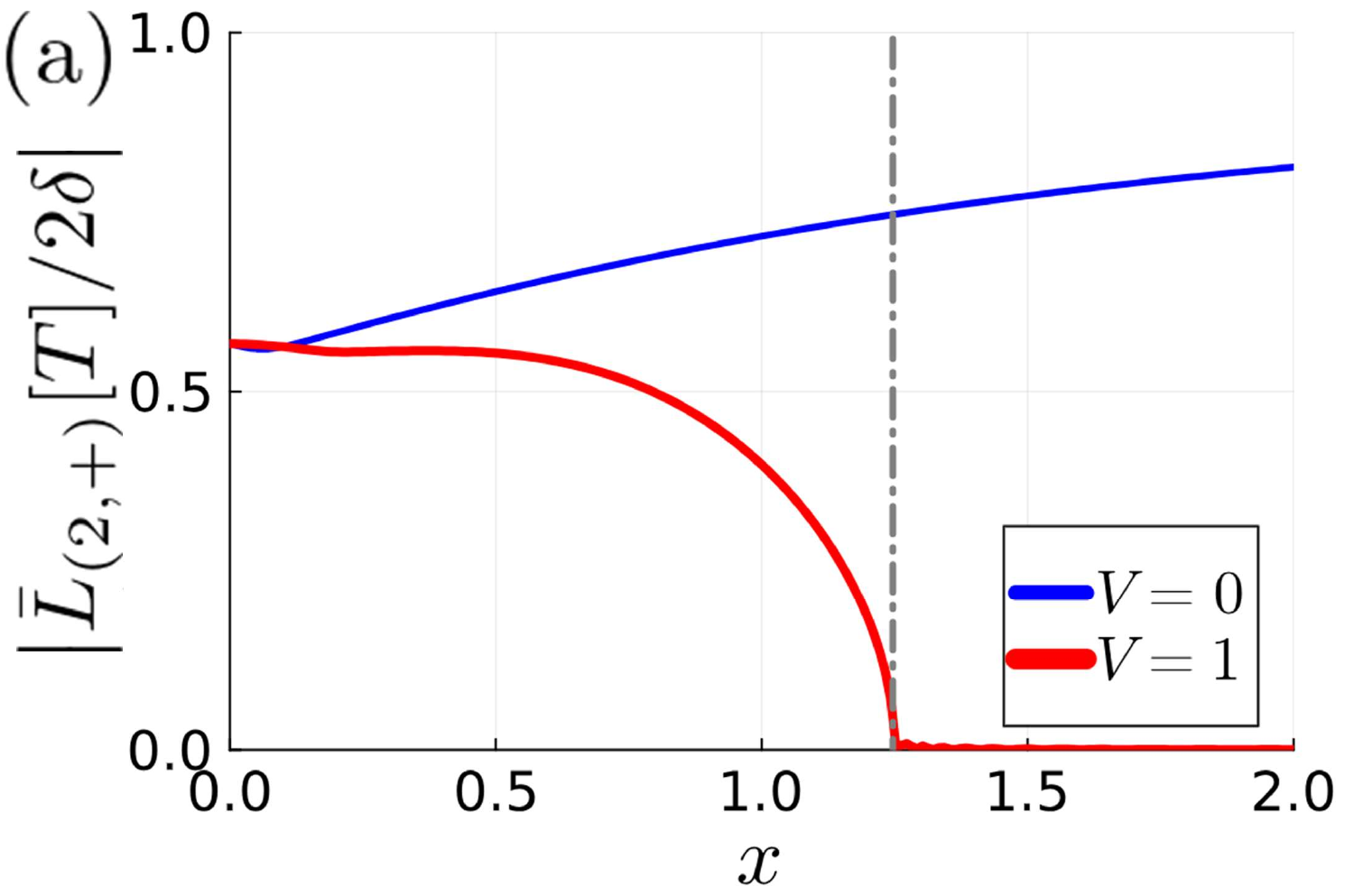}
    \includegraphics[width=0.49\linewidth,height=3.6cm,keepaspectratio]{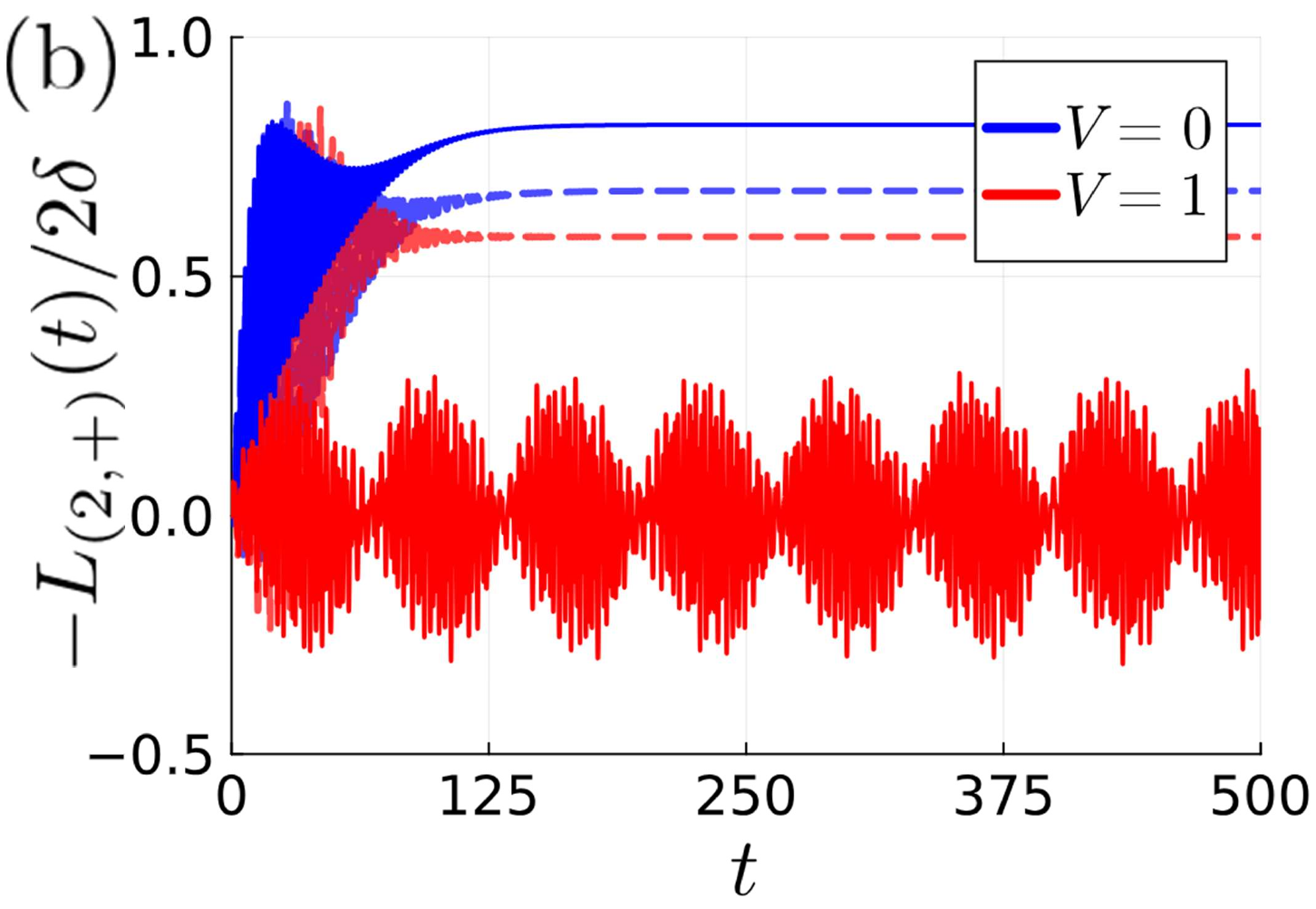}
\end{center}
\end{minipage}
\caption{
(a): Dependence of $|\bar{L}_{(2,+)}[T]/2\delta|$ on $x$ for $(\delta,y,T)=(0.1,1.0,500)$. The gray dashed line indicates $y=1.25$.
(b): Time dependence of $-{L}_{(2,+)}(t)/2\delta$ for $(\delta,y)=(0.1,1.0)$. The dashed and solid lines correspond to $x=0.5$ and $x=1.5$, respectively. In panels (a) and (b), blue and red denote the cases $V=0$ and $V=1$, respectively.
}
\label{fig:IEEL_f_mean_L_Lt}
\end{figure}
%%%%%%%%%%%%%%%%%%%%%%%%%
Figure~\ref{fig:IEEL_f_mean_L_Lt}(a) shows that 
the time-averaged loss rate $\bar{L}_{(2,+)}[T]$ rapidly approaches zero in the vicinity of the interaction-enabled EP2 located at $x\sim1.25$, following the $x$ dependence of the maximum imaginary part of the complex eigenvalues, and becomes $\bar{L}_{(2,+)}[T]\sim 0$ for $x\gtrsim1.25$.
In addition, Fig.~\ref{fig:IEEL_f_mean_L_Lt}(b) shows that the presence of interaction-enabled EL2s is also reflected in the time-dependence of the loss rate.

\end{document}